\documentclass [12pt,a4paper] {article}

\usepackage {graphicx}
\usepackage {xcolor}
\usepackage {amsmath}
\usepackage {amssymb}
\usepackage {slashbox}		%includes backslashbox
\usepackage[affil-it]{authblk}	%for affiliation

\usepackage {ulem}		%contains xout

\begin {document}

\title {Streamer propagation in the atmosphere of Titan and other N$_2$:CH$_4$
mixtures compared to N$_2$:O$_2$ mixtures}

\author {Christoph K\"ohn$^{1}$\footnote{Corresponding author:
koehn@space.dtu.dk}, Sa\v{s}a Dujko$^{2}$, Olivier Chanrion$^{1}$, Torsten Neubert$^{1}$}

\affil{$^{1}$ Technical University of Denmark, National Space Institute (DTU
Space), Elektrovej 328, 2800 Kgs Lyngby, Denmark}
\affil{$^{2}$ Institute of Physics, University of Belgrade, Pregrevica 118,
11080 Belgrade, Serbia}

\date{}

\maketitle

\begin {abstract}
Streamers, thin, ionized plasma channels, form the early stages of lightning
discharges. Here we approach the study of extraterrestrial lightning by
studying the formation and propagation of streamer discharges in various
nitrogen-methane and nitrogen-oxygen mixtures with levels of nitrogen from 20\%
to 98.4\%. We present the friction
force and breakdown fields $E_k$ in various N$_2$:O$_2$
(Earth-like) and N$_2$:CH$_4$ (Titan-like) mixtures.
The strength of the friction force is larger in
N$_2$:CH$_4$ mixtures whereas the breakdown field in mixtures
with methane is half as large as in mixtures with oxygen. We use a 2.5
dimensional Monte Carlo particle-in-cell code with cylindrical symmetry to
simulate the development of electron avalanches from an initial electron-ion
patch in ambient electric fields between $1.5E_k$ and $3E_k$. We compare the electron density, the electric field, the front
velocities as well as the occurrence of avalanche-to-streamer transition between mixtures with methane and with oxygen. Whereas
we observe the formation of streamers in oxygen in all considered
cases, we observe streamer inceptions in methane for small percentages of nitrogen or for large electric fields
only. For large percentages of nitrogen or for small fields, ionization is not
efficient enough to form a streamer
channel within the length of the simulation domain. In oxygen, positive and negative
strea\-mers move faster for small percentages of
nitrogen. In mixtures with methane, electron or streamer fronts move 10-100 times slower than in
mixtures with oxygen; the higher the percentage of methane,
the faster the fronts move.
\end {abstract}

\section {Introduction} \label {intro.sec}
Lightning on Earth is a highly complex phenomenon involving
physical processes on various spatial and temporal scales
\cite{villanueva_1994,ebert_2006,ebert_2008,luque_2012, rakov_2013, silva_2013}
starting from electron avalanches and resulting in the
formation of hot, conducting lightning leaders. The early stages of lightning
discharges are formed by streamers, thin, ionized plasma channels
\cite{loeb_1939,raether_1939,loeb_1940,morrow_1997,ebert_2008,luque_2008,liu_2012,qin_2014}.
Three necessary, yet not sufficient, conditions for the inception of
streamers are the presence of initial
free electrons with a sufficiently high density, a sufficiently high
ambient electric field to accelerate electrons and an ambient gas as
a source of new electrons to sustain the streamer discharge.

In our solar system, lightning activity has been detected on several
planets. While the occurrence of lightning on Venus is still a controversial
question \cite{russell_2007,takahashi_2008,moinelo_2016,perez_2016}, we posses clear evidences of lightning in the atmospheres of
the gas and ice giants. Since Voyager 1, every probe that has approached
Jupiter has imaged lightning flashes from its night side. Lightning has
also been identified on Jupiter by its very-low frequency (VLF) radio
signatures \cite{vasavada_2005,yair_2008}. On Saturn, lightning has been recently detected optically
and by its high-frequency (HF) radio signals \cite{dyudina_2010,dyudina_2013}, known as Saturn
electrostatic discharges (SED) \cite{fischer_2008}. HF radio signals similar to those
observed on Saturn have been detected by the Voyager 2 radio instrument at
Uranus \cite{yair_2008,zarka_1986}. Lightning is also believed to take place at Neptune, based on
the detection of lightning whistler like events observed during the
encounter of Voyager 2 with this planet \cite{gurnett_1990,gibbard_1999},
and on Mars, in dust storms, based on the measurements of higher order moments of the electric field by a detector
installed on NASA's Deep Space Network \cite{renno_2003,ruf_2009} and
supported by simulations by \cite{melnik_1998}. However, more recent
measurements on Mars by Anderson et al. \cite{anderson_2012} and by Gurnett et al.
\cite{gurnett_2010} using the Allen Telescope and the radar receiver of the
Mars Express have not found any signatures of lightning discharges. Hence,
it is still controversial whether lightning exists on Mars.

The existence of atmospheric electric discharges on these planets has been
modelled by numerical simulations and tested by laboratory experiments. 
Using Monte Carlo simulations, Dwyer et al. \cite{dwyer_2006} simulated the runaway
breakdown, which is the discharge initiation through high-energy electrons whose
friction is significantly smaller than for low-energy electrons
\cite{gurevich_1992}, in the atmospheres of
Jupiter and Saturn. They found that the runaway breakdown
field lowered by the presence of hydrometeors is ten times smaller
than the conventional breakdown field and suggest that this might facilitate
lightning inception on these planets. 

Borucki et al. \cite {borucki_1985} experimentally simulated
lightning discharges on Venus, Jupiter and Titan by initiating laser-induced
plasmas in various gas mixtures and found that the emitted spectral lines of these induced
discharges depend significantly on the gas composition. On Jupiter, they
mainly observed spectral lines associated with hydrogen whilst plasmas on Titan and
Venus show spectral lines related to the abundance of the carbon molecules
CH$_4$ and CO. 

Dubrovin et al. \cite{dubrovin_2010} investigated experimentally the inception and
the motion of streamer discharges in CO$_2$:N$_2$ and H$_2$:He as they are
present in the atmospheres of Venus and Jupiter in pressures between 25 and
200 mbar. They discovered that streamer discharges exist in these
atmosphere, yet fainter than on Earth. However, to the best of our
knowledge, there have not been any simulations of streamer discharges and
their inception in non-terrestrial gas compositions modelling the very early stages of lightning
discharges.

Amongst the bodies of our solar system, not only planets, but also some of
their moons shelter an atmosphere, such as Jupiter's satellite Europa
with a pressure of $10^{-11}$ bar \cite{hall_1995} or Saturn's moon Titan. Both satellites are suspected to host aminoacids
\cite{raulin_1995,loison_2015} which
are a necessary condition for the formation of life as we know it
\cite{haldane_1929,oparin_1938,ward_2015}.
Waite Jr. et al. \cite{waite_2007} discussed that cosmic gamma-rays enter Titan's atmosphere
and facilitate the formation of organic compounds, so-called Tholins. Its atmosphere mainly
contains carbonaceous methane and nitrogen which resembles the
atmosphere of the primordial Earth mainly consisting of carbon monoxide,
water, methane and nitrogenous ammonia. Urey and Miller
\cite{miller_1953,miller_1959} mimicked lightning by performing spark discharges in the same gas mixture as of primordial
Earth. They discovered that aminoacids were
formed in their set-up concluding that lightning could be one trigger for the formation of
life. Similarly, Plankensteiner et al. \cite{plankensteiner_2007}
performed discharge experiments in a gas resembling the composition on
Titan and found that possible discharges can produce higher hydrocarbons and
molecules relevant for the formation of amino acids and nucleic acids.

First attempts to detect lightning on Titan were executed in 1980 when Voyager
1 flew by
Saturn and Titan \cite{desch_1990}. Since the thickness of clouds and haze layers prevented the measurements of
optical signatures, the search was extended towards spherics at radio
wavelengths. However, no relevant data was recorded concluding that the
maximum energy of a lightning flash on Titan would be approx. 1 MJ, approx.
1000 times weaker than for lighting flashes on Earth \cite{rakov_2003}.

In 2005, the Huygens Atmospheric Structure Instrument (HASI), designed to
accurately measure atmospheric properties of Titan, measured a resonance at
36 Hz initially linked to the occurrence of lightning
\cite{fulchignoni_2005}. Yet, a subsequent analysis of the provided data
rather suggested that this resonance is an artifact of Titan's interaction
with Saturn's magnetosphere \cite{beghin_2009}.

Successively, Cassini's RPWS (Radio and Plasma Wave Science)
instrument \cite{lammer_2000,fischer_2007,lorenz_2008,fischer_2011} tried to search for
radio emissions as an indicator for lightning on Titan during, in total, 107
flybys. Again, during these missions, no positive results were found;
however, the conclusion was not that lightning does not exist on Titan, but rather
that, if lightning exists, it is too weak to be detected.

Despite the fact that lightning has not been observed directly by cameras
on-board or indirectly by radio instruments during Cassini's observations or
during the descent of the Huygens probe, the atmospheric chemistry suggests
the presence of electrical discharges in the atmosphere of Titan. Titan has
a substantial atmosphere consisting predominantly of nitrogen and methane
with trace amounts of hydrogen, hydrogen cyanide, ethane, propane, acetylene
and other hydrocarbons and nitriles. The presence of the majority of these
hydrocarbons and nitriles in the atmosphere of Titan has been explained in
terms of photochemistry and charged-particle chemistry models with
few exceptions. Perhaps the best known example is ethylene, which is more
abundant in the upper atmosphere of Titan than photochemistry models would
predict \cite{bar-nun_1979,borucki_1988}. Borucki et al. \cite{borucki_1988} argued that the excess of ethylene in the
upper Titan's atmosphere could be explained by upward diffusion of ethylene
produced in the lower parts of the atmosphere by lightning. Likewise, the
presence of acetylene and hydrogen cyanide with the relatively high
abundances of $(1.9\pm 0.2)\times10^{-6}$ and $(1.5\pm0.2)\times10^{-7}$ \cite{taylor_1998}, respectively, could
also be explained by the lightning induced chemistry.  

Whereas Earth's surface conductivity is approximately $10^{-14}$ S m$^{-1}$ \cite{wahlin_1994}, Molina-Cuberos et al.
\cite{molina_2001} approximated Titan's surface conductivity to range between
$10^{-15}$ S m$^{-1}$ and $10^{-10}$ S m$^{-1}$, thus sufficiently large to
allow the generation of cloud-to-ground lightning flashes.
Tokano et al. give an extensive overview of available cloud
convection and charging models as well as particle charging mechanisms (see
\cite{tokano_2001} and references therein). They apply a one-dimensional
time-dependent thundercloud model and state that negative space charges
resulting from the attachment of electrons to clouds, can temporally create
electric fields of up to 2 MV m$^{-1}$ which is sufficient to initiate 20 km
long negative cloud-to-ground lightning. 

We here strive to answer the question whether streamer discharges, the
pre-cursors of lightning, exist in Titan's atmosphere. Therefore, we
perform Monte Carlo
particle-in-cell simulations of electron avalanches in mixtures of
N$_2$:CH$_4$ with different percentages of nitrogen in various electric fields and determine for which
conditions the electron avalanches transition into streamer discharges.
We also run simulations in N$_2$:O$_2$ mixtures with the same percentage of
nitrogen and compare results for methane and for oxygen.

In section \ref{model.sec} we briefly discuss the atmospheric profile of
Titan and the set-up of our simulations. Additionally, we discuss the
friction forces as well as the electric breakdown fields in different N$_2$:O$_2$ and N$_2$:CH$_4$
mixtures. In section \ref{results.sec} we discuss the temporal evolution of the electron densities, the front
velocities and the resulting electric field and compare results
in mixtures with methane and with oxygen. We discuss when
avalanche-to-streamer transitions are feasible and relate our results to the
friction force and thus to the electron energy distribution. In section
\ref{concl.sec} we summarize our results and discuss whether lightning is
possible to occur on Titan. Finally, we give an outlook on future
research activities.

\section {Modelling and properties of Titan's atmosphere} \label {model.sec}

\subsection {Set-up of the model} \label{setup.sec}
We simulate the development of electron avalanches
and, if existent, of subsequent streamers with a 2.5D Monte Carlo
Particle-In-Cell code with cylindrical symmetry with two spatial coordinates $(r,z)$ and three
coordinates ($v_r,v_{\theta},v_z$) in velocity space for each individual
(super)electron \cite{chanrion_2008,chanrion_2010}. We perform
simulations in mixtures of N$_2$:O$_2$ and of N$_2$:CH$_4$ with
different percentages $\kappa$ of N$_2$. Thus, Earth's atmosphere is
determined by $\kappa=0.8$ with its nitrogen-oxygen mixture and Titan's
atmosphere by $\kappa=0.984$ with its nitrogen-methane mixture. For
a better comparison of the results in different gas mixtures, 
we use a gas density of $2.9\cdot 10^{25}$ m$^{-3}$ in all mixtures
(see section \ref{atmos.sec}).

The collision of electrons with nitrogen and oxygen molecules, cross sections as well as their
implementation have already been studied carefully in various publications (e.g.
\cite{gurevich_1961,phelps_1985,crompton_1994,moss_2006,chanrion_2008,dujko_2011,li_2012,koehn_2014,koehn_2015}). We
have reviewed the collisions of electrons with methane molecules and the resulting friction
force in section \ref{cross.sec}.

Accounting for space charge effects, we solve the Poisson equation in the
simulation domain with dimensions
$(L_r,L_z)$=(1.25, 14) mm and with 150 grid points in $r-$ and 1200 grid points
in $z-$direction \cite{koehn_2017}.  As boundary conditions, we use Neumann conditions
$\partial\phi/\partial r=0$ for $r=0,L_r$, and Dirichlet conditions for the
electric potential $\phi(r,0)=0$ and $\phi(r,L_z)=\phi_{max}=E_{amb}\cdot
L_z$ where $E_{amb}$ is the ambient electric field pointing downwards. We
have performed simulations in ambient fields of 1.5, 2 and 3 times the
classical breakdown field $E_k$. %where $E_k$ depends on the gas mixture.

We initiate all simulations with a charge neutral electron-ion patch at the
center of the simulation domain. The initial electron density is given by
the Gaussian $n_e(r,z,t=0)=n_{e,0}
\exp\left(-\left(r^2+(z-z_0)^2\right)/\lambda^2\right)$
with $n_{e,0}=10^{20}$ m$^{-3}$ as in \cite{arrayas_2002,liu_2004,koehn_2017} and $\lambda=0.2$ mm centered at $z_0=7$ mm.

\subsection {Titan's atmospheric composition} \label {atmos.sec}
An overview of the abundances of the constituents of Titan's atmosphere
is given in \cite{niemann_2005}.
Titan's atmosphere is composed mainly of nitrogen and of methane
where the percentage of methane varies from approx. 5\%
at ground up to approx. 1.4\% at 140 km altitude. In order to perform reliable simulations, it is, however, not sufficient to know the gas composition, but we also need to
specify the correct number density of ambient gas
molecules. Lindal et al. and McKay et al. \cite{lindal_1983,mckay_1989} give an extended overview of
Titan's temperature and pressure profile as a function of altitude.
Further measurements of the temperature and pressure were
later performed by the Huygens Atmospheric Structure Instrument (HASI)
descending towards Titan's surface \cite{fulchignoni_2005}.

On Titan, clouds form between 20 km and 35 km altitude
\cite{barth_2007,griffith_2009} with
the pressure varying between 
approximately 0.1 bar and 0.6 bar, the temperature varying between
70 K and 75 K and the level of methane varying between 1.6\% and 2.0\%
\cite{mueller-wodarg_2014}. Hence, we here choose
1.6\% of methane as well as $p=0.3$ bar and $T=75$ K
yielding a gas density of $n_{Titan}=p/(k_B T)\approx 2.9\cdot 10^{25}$
m$^{-3}$ using the ideal gas law. On Earth a density of $2.9\cdot
10^{25}$ m$^{-3}$ corresponds to a pressure of approx. 1.2 bar at 300 K or
approx. 1 bar at 250 K (equivalent to 10 km altitude).

\subsection {Cross sections and
friction forces for electrons in N$_2$:O$_2$ and N$_2$:CH$_4$ mixture} \label{cross.sec}
%%%%%%%%%%%%%%%%%%%%%%%FIG. 1%%%%%%%%%%%%%%%%%%%%%%%%%%%%%%%%%%%%%%%%%
\begin {figure}
\begin {center}
\begin {tabular}{cc}
\includegraphics [scale=0.28] {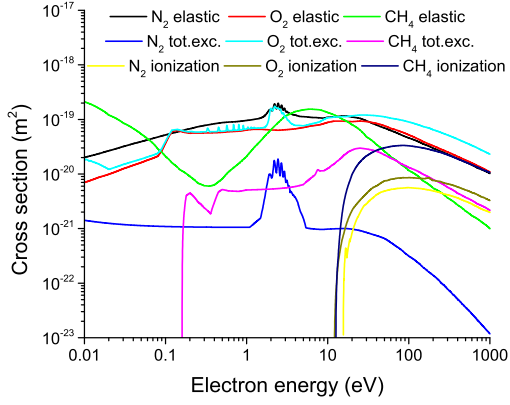} &
\includegraphics [scale=0.28] {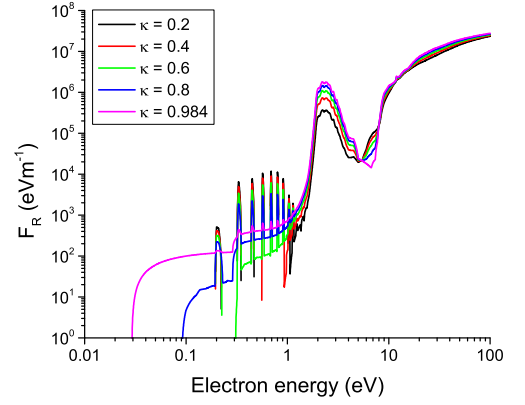} \\
a) $\sigma_{i,j}$ & b) $F_R$, N$_2$:O$_2$\\
\includegraphics [scale=0.28] {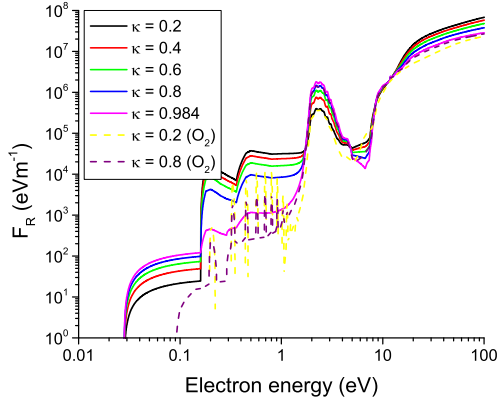}\\
c) $F_R$, N$_2$:CH$_4$
\end {tabular}
\end {center}
\caption {a) Cross sections for electron scattering in N$_2$, O$_2$ and
CH$_4$. We show cross sections for momentum transfer in elastic collisions, total
cross sections for electronic excitation (including dissociation into
neutral fragments) and cross sections for electron-impact ionization. b,c) The friction force in N$_2$:O$_2$ (b) and in N$_2$:CH$_4$
(c) as a function of
electron energy for different percentages $\kappa$ of molecular nitrogen. For
comparison, the dashed yellow and purple lines in c) show the friction force for a N$_2$:O$_2$
mixture with 20\% and 80\% nitrogen.} \label{cross.fig}
\end {figure}
%%%%%%%%%%%%%%%%%%%%%%%%%%%%%%%%%%%%%%%%%%%%%%%%%%%%%%%%%%%%%%%%%%%%%%

Figure \ref{cross.fig} a) shows the cross sections $\sigma_{i,j}$ for
different collision processes $j$ of electrons at different species $i$. As for oxygen, electrons scatter off methane elastically, excite, ionize
or attach to it. The most frequent process below approximately 50 eV is
elastic scattering followed by excitations
\cite{sasic_2004}. Since the ionization threshold energy is $E_b\approx 12.6$ eV, the
ionization cross section starts to increase for energies above and becomes
predominant above approximately 50 eV.

However, due to the lack of reliable data, we do not have differential cross sections
for elastic and
inelastic collisions, including those for ionization. However, as we
discuss below, most of the electrons in the N$_2$:CH$_4$ simulations have energies below
50 eV. For such low energies, the collision dynamics is well described by the
approximation of isotropic scattering. The isotropy of scattering may be
assumed due to the fact that numerous elastic collisions generally randomize directions of
electron motion and hence one could not see any significant effects of
introducing the anisotropy of scattering. The determination of the
momentum transfer cross section is usually performed under the
assumption of anisotropy, so the cross section has this assumption
inherently built into it. Of course, for higher electron energies the
approximation of isotropic scattering is no more valid and the use of
differential cross sections for electron scattering is mandatory (see
for example a detailed discussion in \cite{li_2009}). The energy $W$ of
secondary ionization electrons is determined uniformly randomly in $[0,E_{in}-E_b]$ where
$E_{in}$ is the kinetic energy of the incident electron.
Subsequently, we apply the conservation of energy and momentum and determine
\cite{koehn_2014,celestin_2010} the scattering angle
\begin {eqnarray}
\cos\Theta_{sca}=\sqrt{\frac{(E_{in}-W)(E_{in}+2m_ec^2)}{E_{in}(E_{in}-W+2m_ec^2)}}
\label {angle.1}
\end {eqnarray}
of the incident electron and the emission angle
\begin {eqnarray}
\cos\Theta_e=\sqrt{\frac{W(E_{in}+2m_ec^2)}{E_{in}(W+2m_ec^2)}}.
\label{angle.2}
\end {eqnarray}
of the liberated electron with $m_e$ being the electron's rest mass and $c$
the speed of light. Note that these relativistic equations are valid also
for non-relativistic electron energies.

Figure \ref{cross.fig} b,c) show the friction force
\begin {eqnarray}
F_R(E)=\sum\limits_{i,j} n_i \sigma_{i,j}(E) \Delta E_{i,j}, \label {friction}
\end {eqnarray}
where $n_i$ is the partial density [m$^{-3}$] of N$_2$, O$_2$ or CH$_4$,
$\sigma_{i,j}$ is the total cross section for collision process $j$
of electrons at molecule species $i$ and $\Delta E_{i,j}$ the respective
energy loss of electrons.
The sum is taken over all involved molecule species $i$ and over
all involved inelastic collisions $j$. Note that excitations and ionization are the only processes below 100
eV contributing to the friction where the energy loss resulting from
excitations is the threshold energy of the collision and the energy loss for
ionization is $1/2(E_{in}-E_b)$. Panel b) shows the friction of electrons in different
mixtures of N$_2$:O$_2$. The overall shape is similar irrespective of
the percentage $\kappa$ of nitrogen. However, the strength of excitational losses
for energies below 1 eV increases with decreasing $\kappa$ since the cross section and subsequently
the contribution of excitations is higher for oxygen than for nitrogen.
On the contrary, the contribution of excitations above 1 eV increases with increasing
percentage of nitrogen. For $E\gtrsim 1$ eV, there is a resonance in the
cross section of the vibrational states of excited nitrogen with threshold
energies of approximately 2 eV (see for example Tab. 1 in \cite{moss_2006}).
For energies above approximately 10 eV,
the friction forces for different $\kappa$ align because the total
ionization cross section and the corresponding energy loss are comparable
for nitrogen and oxygen.

Panel c) shows the friction force for different N$_2$:CH$_4$ mixtures. For
comparison, the yellow and purple lines additionally show the friction force
for N$_2$:O$_2$ with 20\% and 80\% nitrogen.  Whilst the overall shape of
the friction force is different than for N$_2$:O$_2$, we observe the same
dependency on $\kappa$: Since the cross section for exciting methane for
electron energies below 1 eV is larger than for nitrogen, the friction force
increases for an increasing percentage of methane.  Because of the resonance
of the cross section for exciting nitrogen above 1 eV, the friction
increases with $\kappa$ for energies between 1 eV and 10 eV.  Comparing the
friction force in N$_2$:O$_2$ mixtures with the friction force in
N$_2$:CH$_4$ mixtures reveals that, for fixed $\kappa$, the strength of the
friction force above approximately 1 eV is similar for mixtures with oxygen
and with methane, hence the energy loss of electrons above 1 eV is
comparable.  However, it is noticeable that for energies below 1 eV and
fixed $\kappa$, the friction
in N$_2$:CH$_4$ mixtures is approximately one to two orders of magnitude higher than in mixtures with
oxygen.

\subsection {Photoionization} \label {photo.sec}
In oxygen-nitrogen mixtures, one additional process contributing to the
evolution of streamer discharges is photoionization: Electrons can excite
nitrogen which subsequently emits UV photons. In return, these UV photons
ionize oxygen and thus deliver a new source of electrons
\cite{zheleznyak_1982,luque_2007,bourdon_2007,wormeester_2010,koehn_2017}. Although
this process is not crucial for the development for negative fronts, it
supports their motion (see e.g. a discussion in \cite{luque_2007,koehn_2017}). For positive streamer fronts moving towards the
cathode, however, photoionization is one of the key drivers next to
background ionization \cite{panchesnyi_2005,nijdam_2011}.

In oxygen, we use the model of Zheleznyak et al. \cite{zheleznyak_1982,
chanrion_2008,koehn_2017} which relates the number of UV photons with energies between
12.10 eV and 12.65 eV to the number of electron impact ionization.  This is
the energy range where photons do not interact with nitrogen and
predominantly ionize molecular oxygen.  Beyond 12.65 eV, photons mainly
excite nitrogen and do not contribute to the ionization of oxygen anymore
\cite{liu_2012, carter_1972}. The ionization energy of molecular nitrogen is 15.6 eV, thus larger than the
upper limit for UV photoionization. Note that the threshold energy for the
ionization of oxygen is 12.1 eV, thus on the lower limit of the energy of
considered UV photons.

Since the ionization energy of methane is 12.6 eV, photoionization by UV
photons still contributes to the development of electron avalanches or
streamers in N$_2$:CH$_4$, even though less efficiently. We thus implement the model of
Zheleznyak et al., but evaluate the UV photoionization process randomly in (12.65-12.6)/(12.65-12.1)=9\%
of all photoionization events in the model by Zheleznyak et
al. only which is the ratio of the energy intervals of UV
photoionization in methane and in oxygen.

\subsection {The electric breakdown field} \label {elec.sec}
%%%%%%%%%%%%%%%%%%%%%%%FIG. 2%%%%%%%%%%%%%%%%%%%%%%%%%%%%%%%%%%%%%%%%%
\begin {figure}
\includegraphics [scale=0.7] {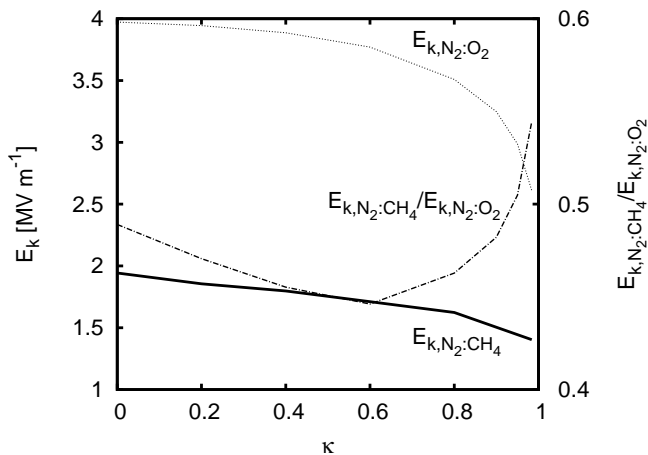}
\caption {The breakdown field $E_k$ as a function of the percentage $\kappa$ of
nitrogen in N$_2$:O$_2$ mixtures and in
N$_2$:CH$_4$ mixtures with a gas density of $2.9\cdot 10^{25}$ m$^{-3}$. The plot also shows the ratio of the breakdown fields
between these two different mixtures (right $y$-axis).} \label{breakdown.fig}
\end {figure}
%%%%%%%%%%%%%%%%%%%%%%%%%%%%%%%%%%%%%%%%%%%%%%%%%%%%%%%%%%%%%%%%%%%%%%

During the motion of electrons through ambient gas, electrons undergo two competing
processes: First, they gain energy through the background electric field and they lose energy
through inelastic collisions with the ambient gas molecules. Second, amongst these inelastic
collisions, electrons are capable of ionizing the ambient gas and multiply
the total electron number as long as the primary electron's energy is above the
ionization threshold energy. As a competing process, electrons attach to molecules
reducing the total electron number. These two processes compete with each other and severely depend on the ambient electric
field \cite{raizer_1991}. The breakdown (or equilibrium) field $E_k$ is defined as the electric field when these two
processes are in equilibrium, i.e. the rate of attachment equals the rate
of ionization. Thus, in order to initiate and sustain an electron avalanche
and eventually transition into a streamer, electric fields at least
above the breakdown field are required such that there is a sufficient
number of ionization events.

In an accompanying paper, we have determined swarm parameters like
the ionization coefficients and the breakdown field for various
mixtures of N$_2$:O$_2$ and N$_2$:CH$_4$ by applying a
multi-term approach for solving the Boltzmann equation \cite{bosnjakovic_2018}. Figure \ref{breakdown.fig} shows
the breakdown field as a function of the percentage $\kappa$ of nitrogen for
a density of $2.9\cdot 10^{25}$ m$^{-3}$ as
well as the ratio between the breakdown fields in N$_2$:O$_2$ and
N$_2$:CH$_4$. It shows that in both gas mixtures, the breakdown field slightly decreases
as a function of $\kappa$. It also shows that the ratio of the breakdown
field in N$_2$:O$_2$ and in N$_2$:CH$_4$ is approximately 0.5 irrespective
of $\kappa$ although increasing for large $\kappa$. Whilst the electric field strength for breakdown varies between
3 and 4 MV m$^{-1}$ in N$_2$:O$_2$, it amounts to only 1.5-2 MV m$^{-1}$ in
N$_2$:CH$_4$ mixtures.

\section {Results} \label {results.sec}

\subsection {Temporal evolution of electron avalanches and strea\-mers for different
percentages of nitrogen}

%%%%%%%%%%%%%%%%%%%%%%%FIG. 3%%%%%%%%%%%%%%%%%%%%%%%%%%%%%%%%%%%%%%%%%
\begin {figure}
\begin {center}
\begin {tabular}{ccc}
\includegraphics[scale=0.30] {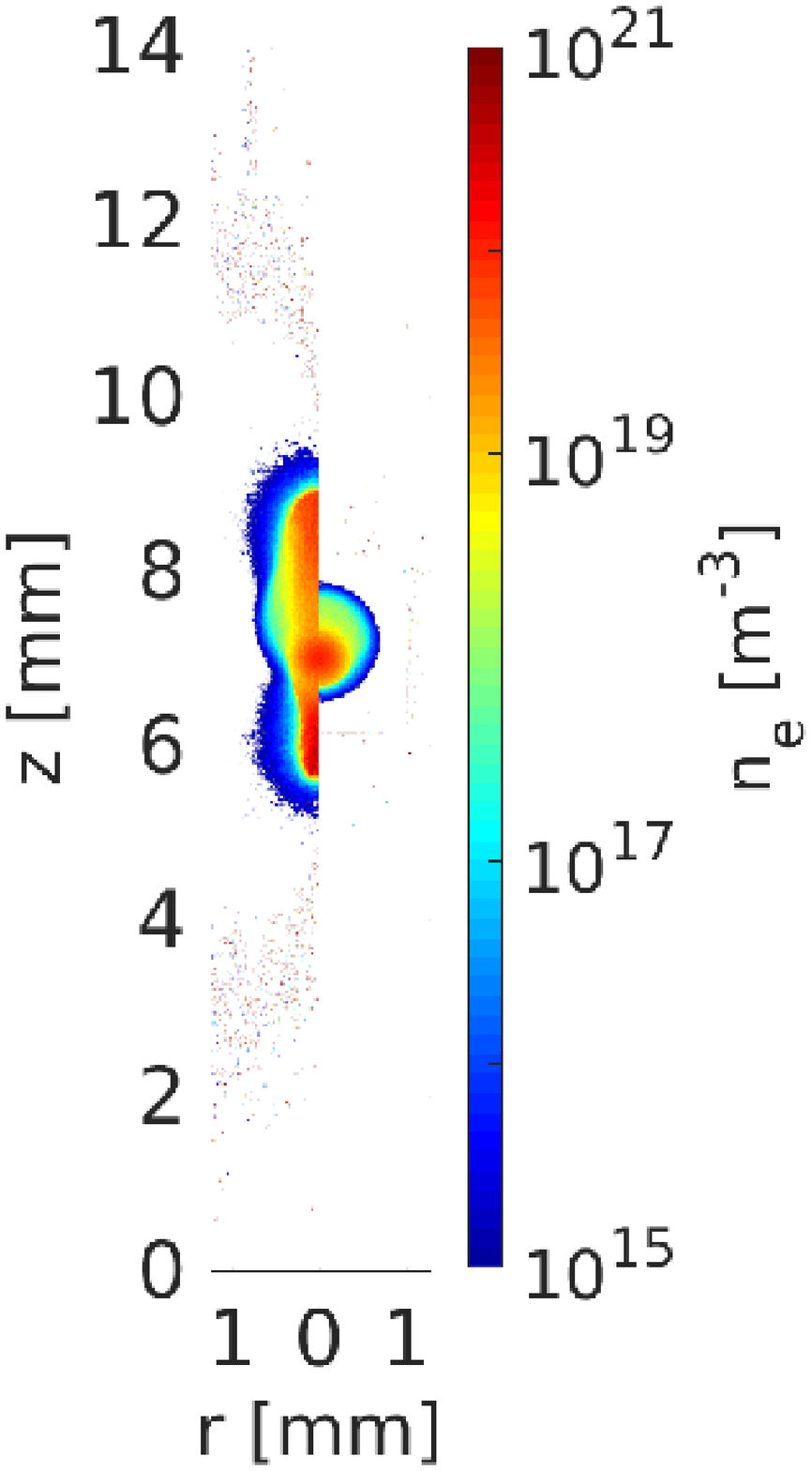} &
\includegraphics[scale=0.30] {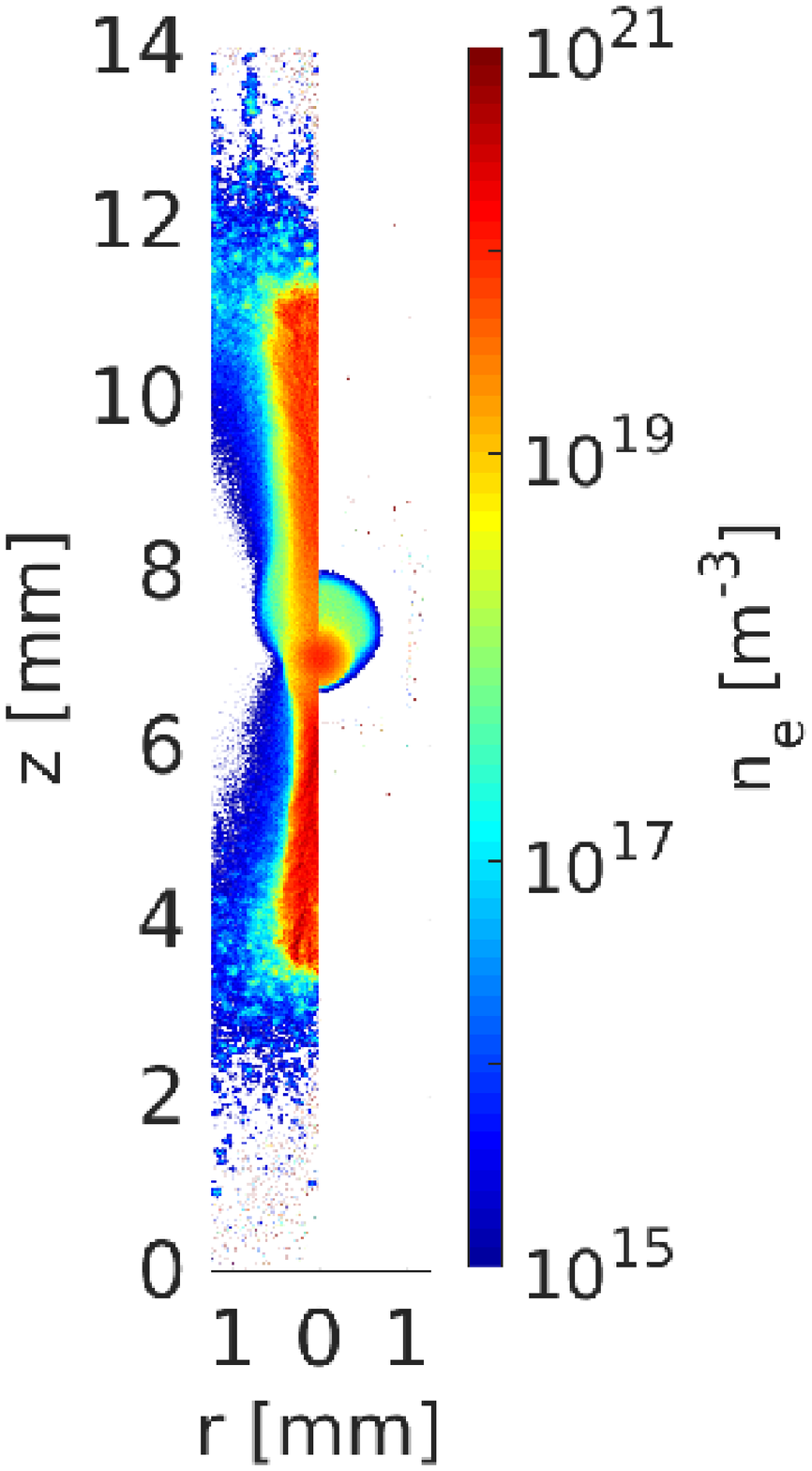} &
\includegraphics[scale=0.30] {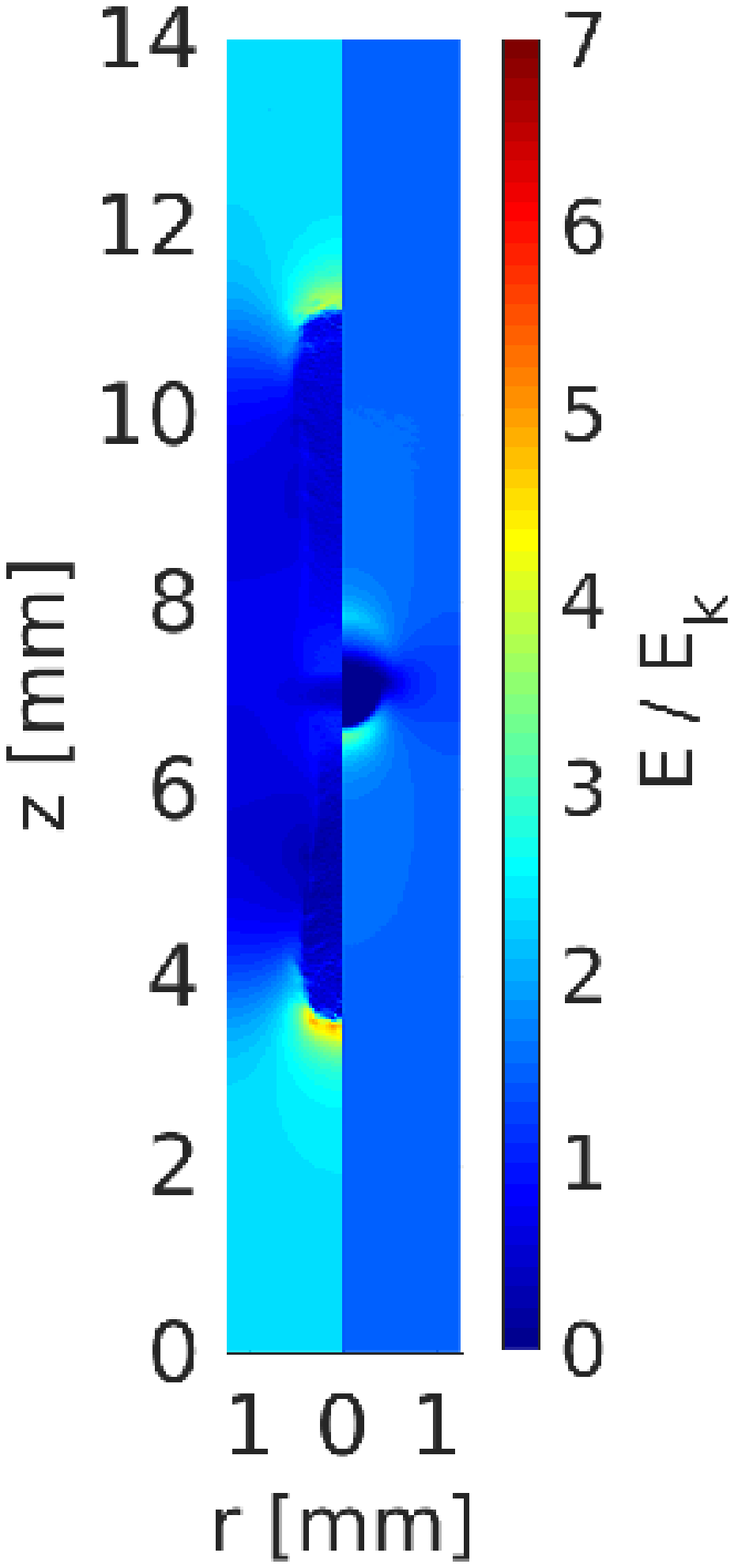}\\
a) $n_e$, $t=1.51$ ns & b) $n_e$, $t=2.57$ ns & c) $E/E_k$, $t=2.57$ ns\\
\includegraphics[scale=0.30] {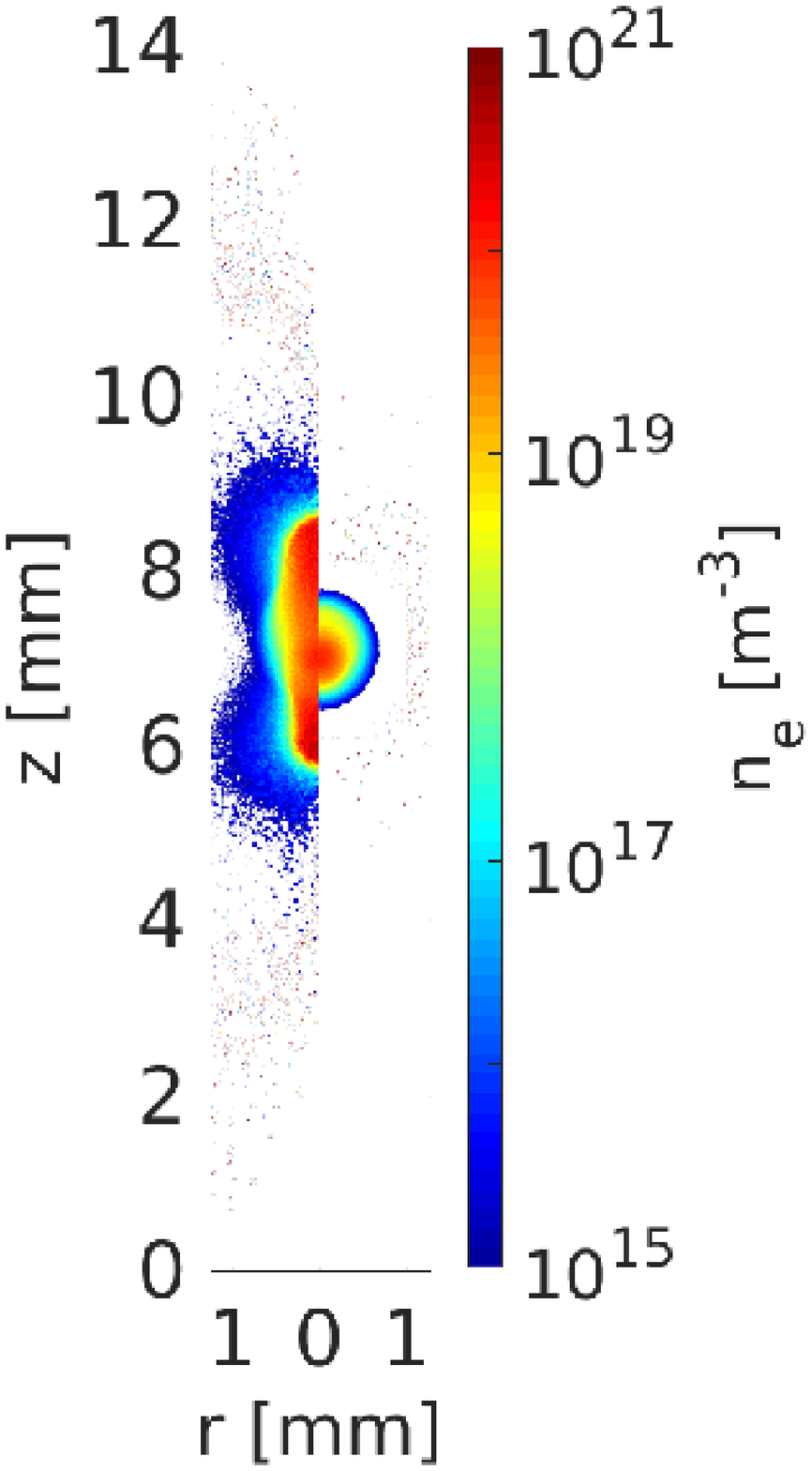} &
\includegraphics[scale=0.30] {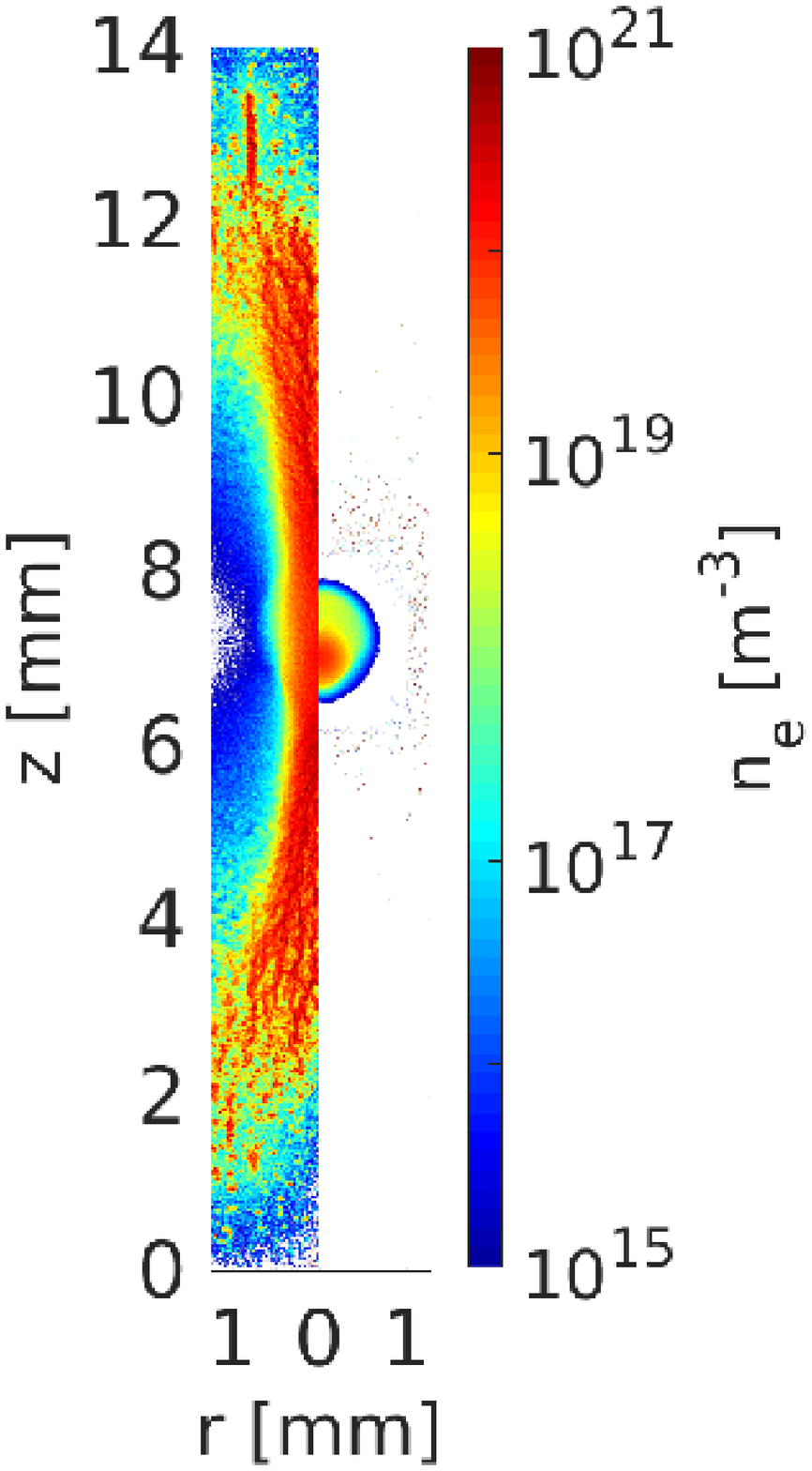} &
\includegraphics[scale=0.30] {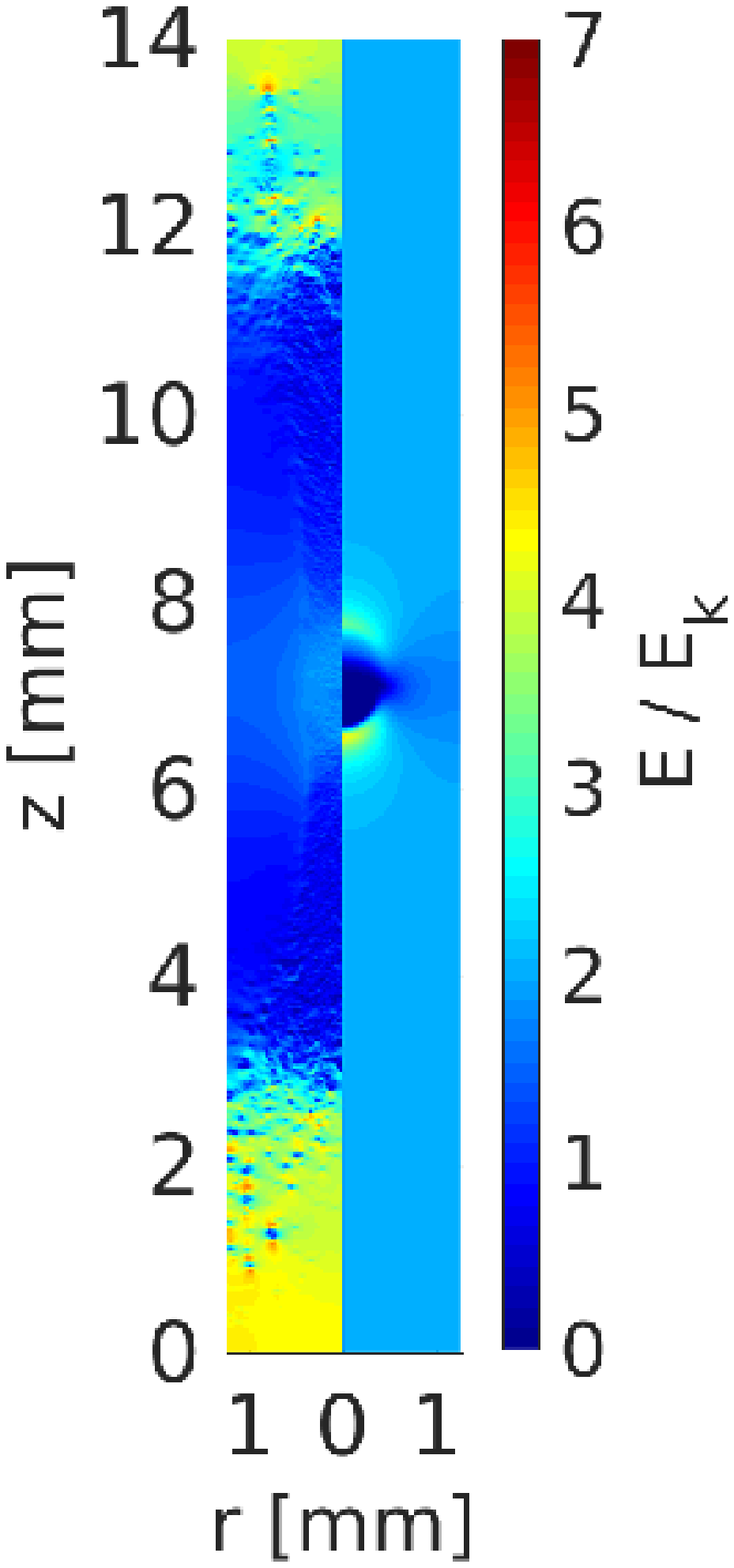} \\
d) $n_e$, $t=0.66$ ns & e) $n_e$, $t=1.33$ ns &
f) $E/E_k$, $t=1.33$ ns\\
\includegraphics[scale=0.30] {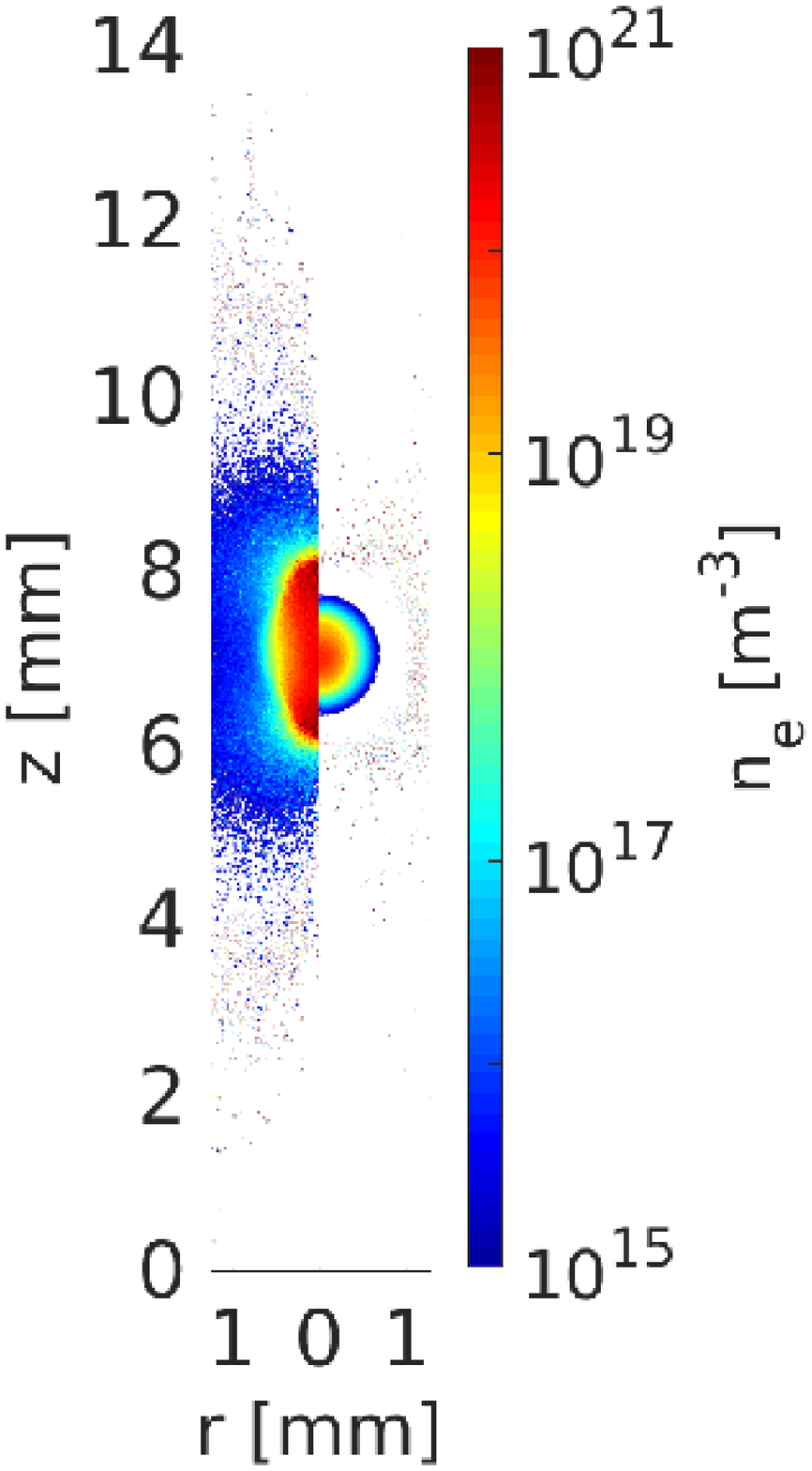} &
\includegraphics[scale=0.30] {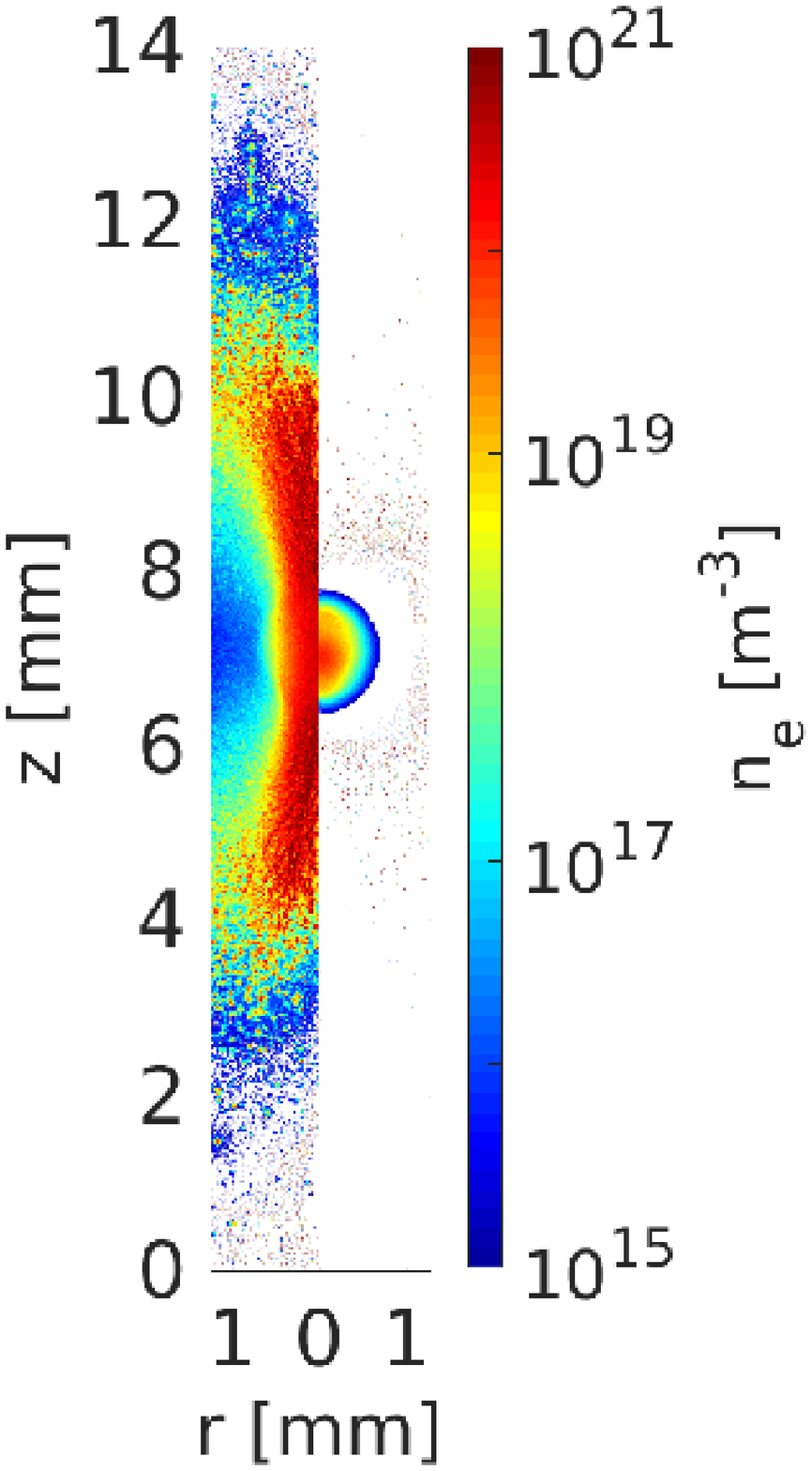} &
\includegraphics[scale=0.30] {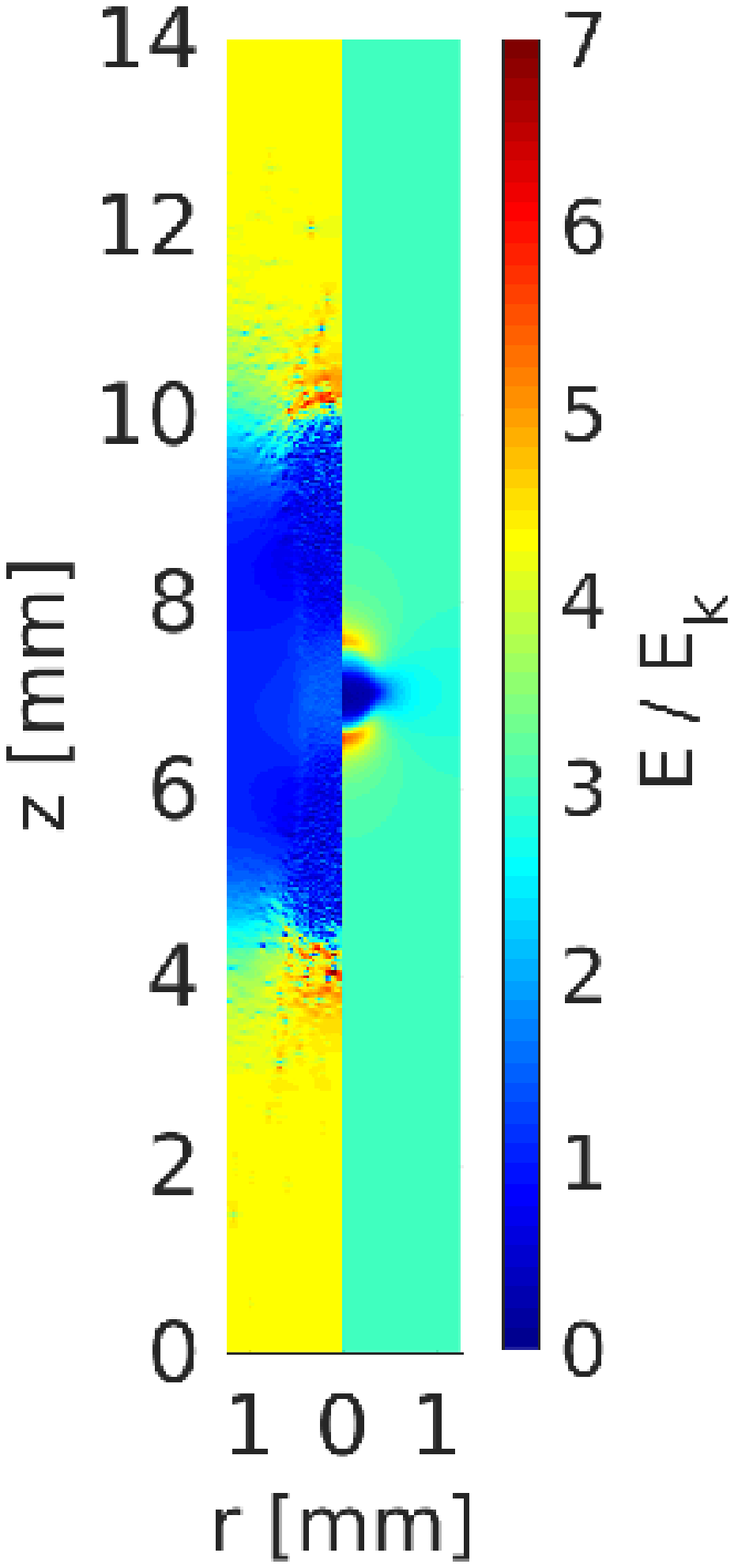} \\
g) $n_e$, $t=0.20$ ns & h) $n_e$, $t=0.40$ ns &
i) $E/E_k$, $t=0.40$ ns\\
\end {tabular}
\end {center}
\caption {The electron density $n_e$ and the electric field $E/E_k$ in the gas
mixtures N$_2$:O$_2$ and N$_2$:CH$_4$ with 80\% nitrogen in an ambient field
of $1.5E_k$ (first row), $2E_k$ (second row) and $3E_k$ (third row). Columns 1 and 2
compare the electron density in mixtures with 20\% oxygen (left half of each panel) with
the density in mixtures with 20\% methane (right half). Column 3 compares the electric
field.} \label{dens_1.fig}
\end {figure}
%%%%%%%%%%%%%%%%%%%%%%%%%%%%%%%%%%%%%%%%%%%%%%%%%%%%%%%%%%%%%%%%%%%%%%

%%%%%%%%%%%%%%%%%%%%%%%NEW FIG. 4%%%%%%%%%%%%%%%%%%%%%%%%%%%%%%%%%%%%%
\begin {figure}
\begin {center}
\begin {tabular}{ccc}
\includegraphics[scale=0.35] {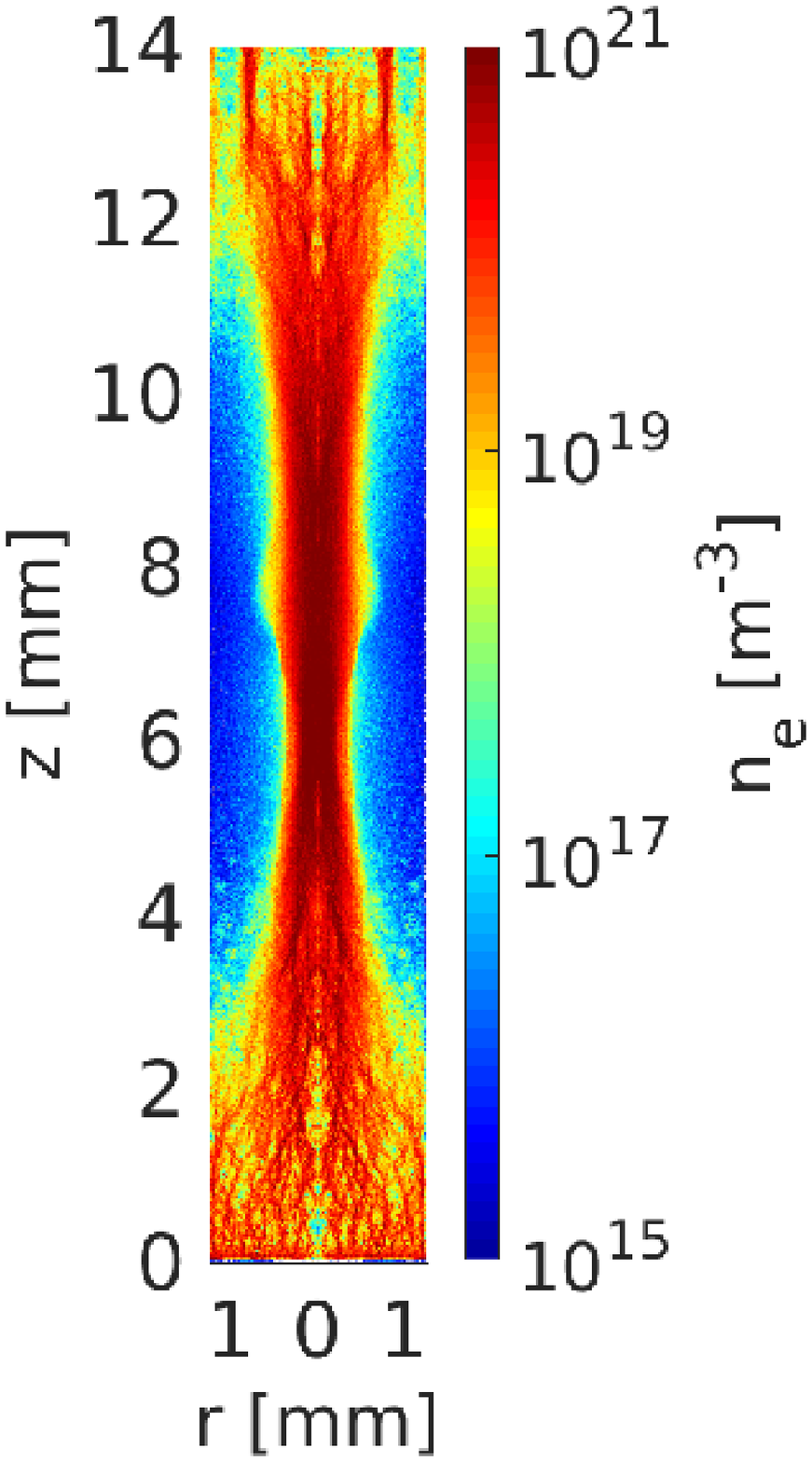} &
\includegraphics[scale=0.35] {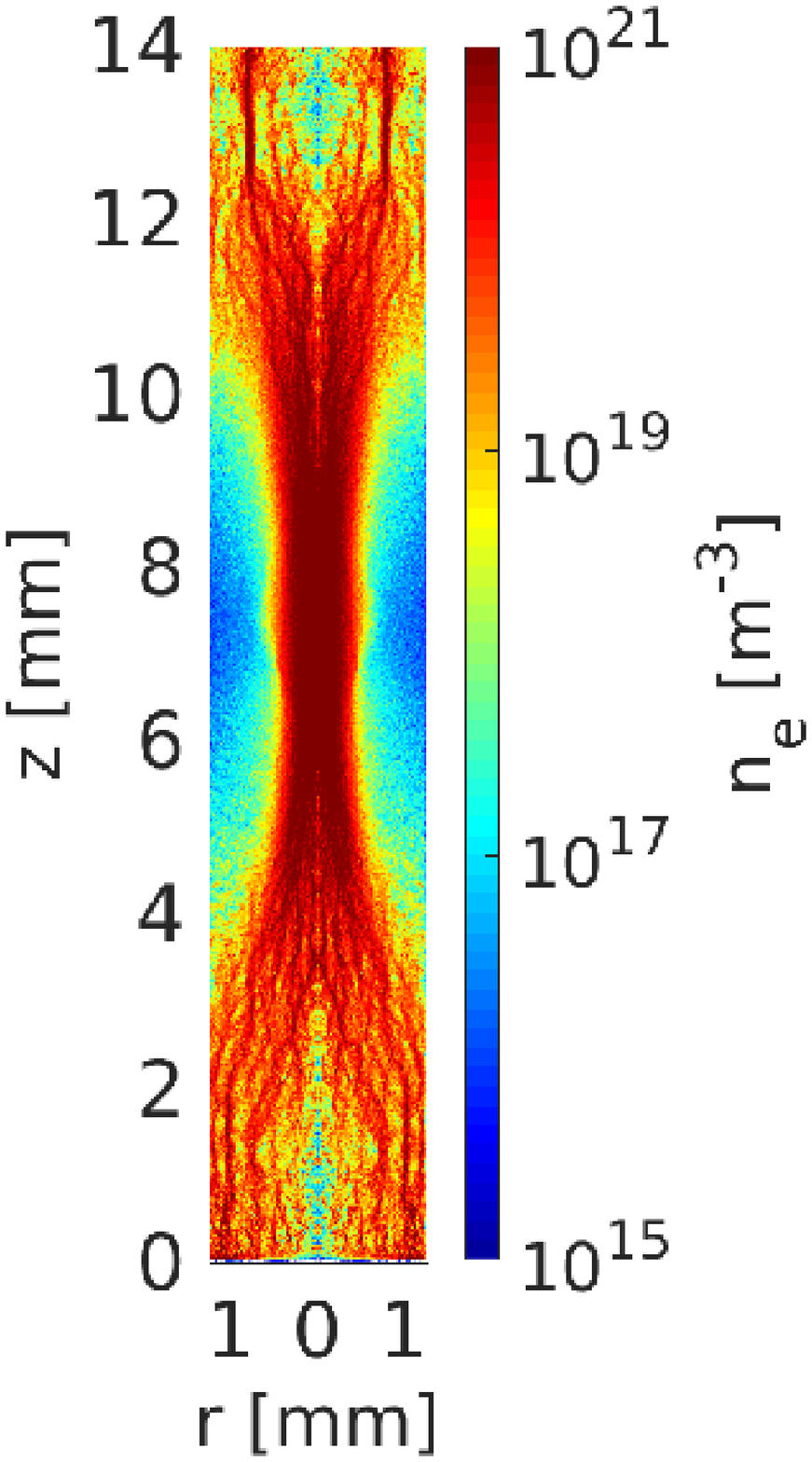}  &
\includegraphics[scale=0.35] {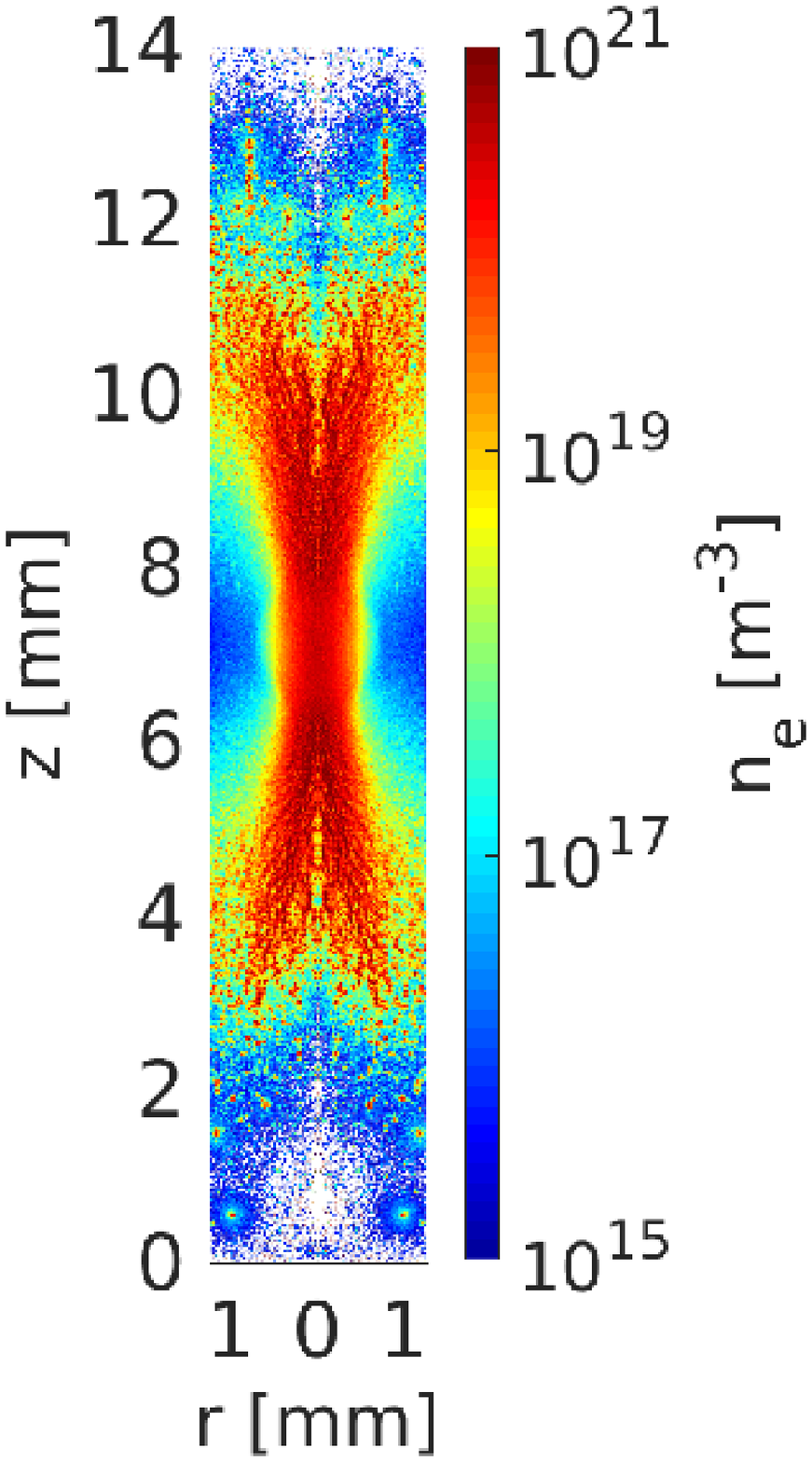} \\
a) $t=3.21$ ns & b) $t=1.49$ ns & c) $t=0.45$ ns \\
\includegraphics[scale=0.35] {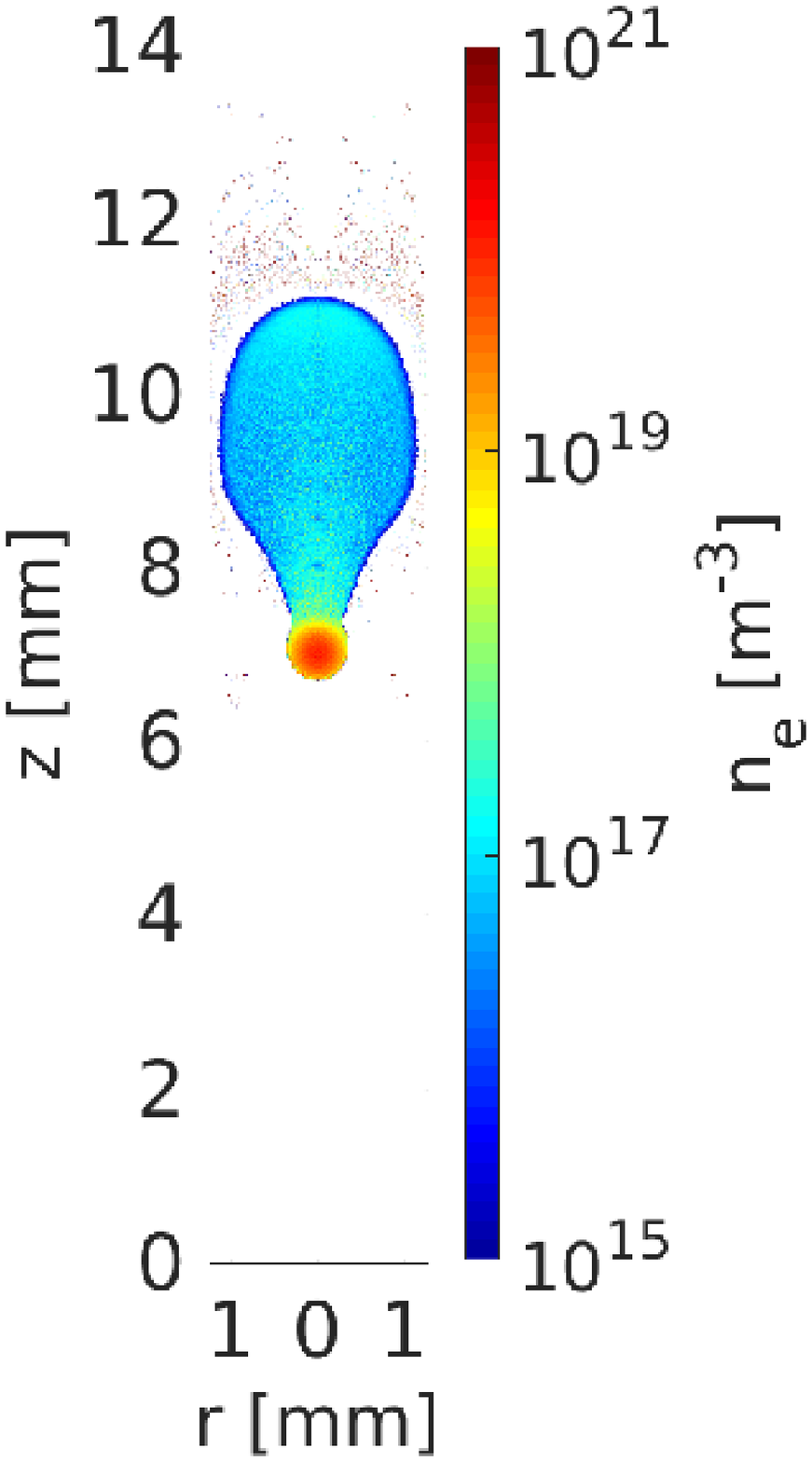} &
\includegraphics[scale=0.35] {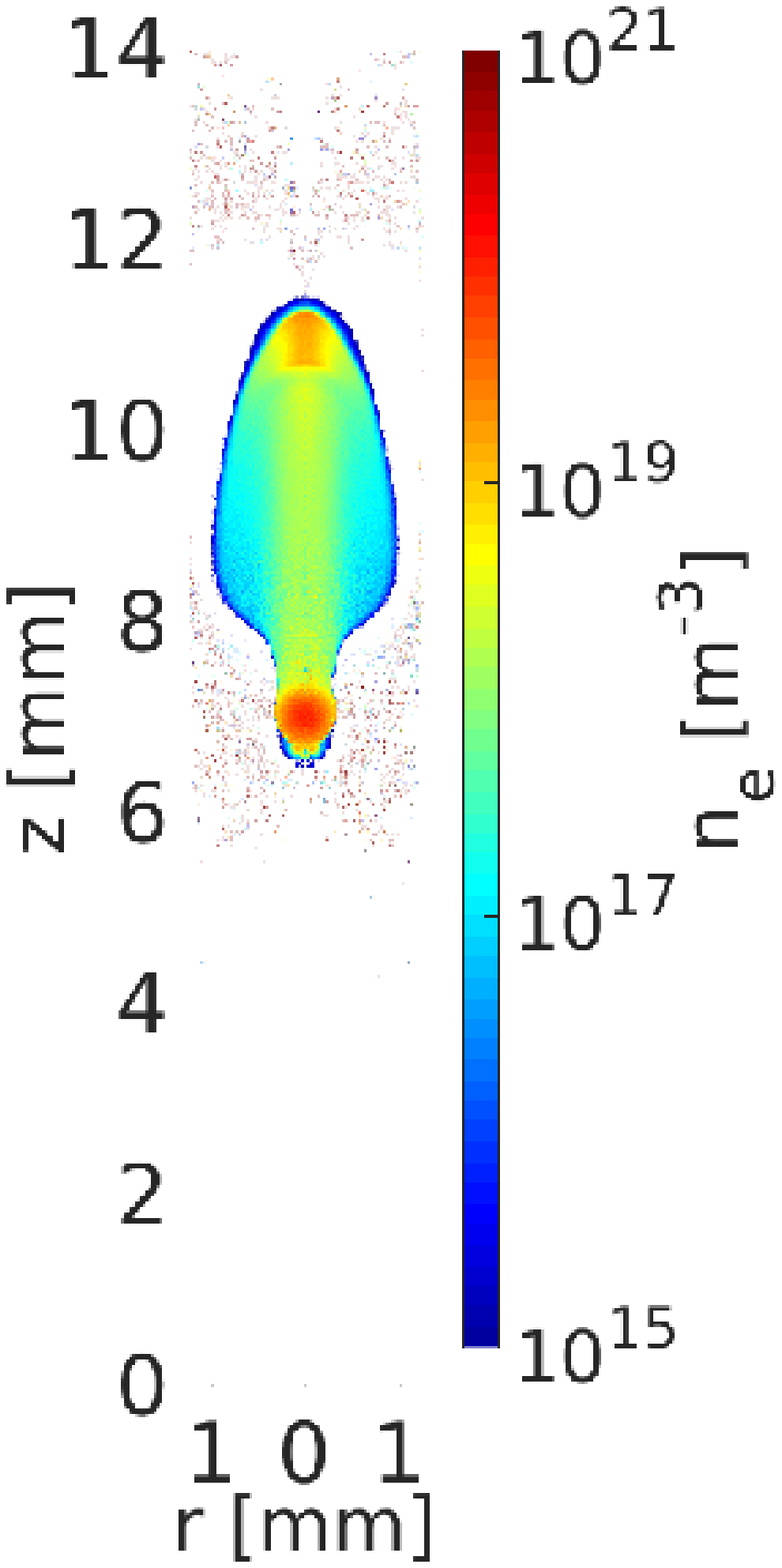}  &
\includegraphics[scale=0.35] {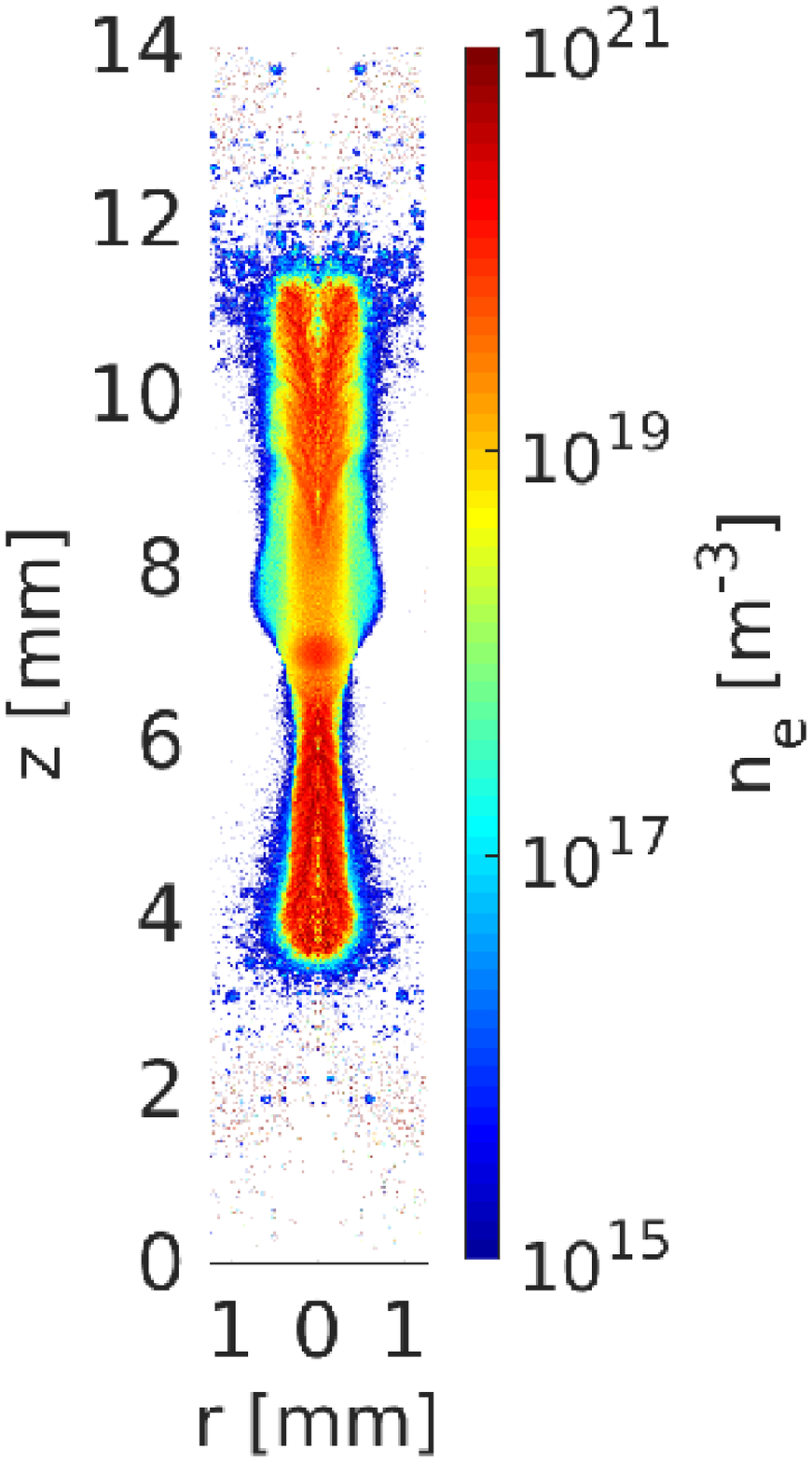} \\
d) $t=28.54$ ns & e) $t=15.92$ ns & f) $t=4.28$ ns\\
$E=1.5E_k$ & $E=2E_k$ & $E=3E_k$\\
\end {tabular}
\end {center}
\caption {The electron density $n_e$ in N$_2$:O$_2$ (first row) and in
N$_2$:CH$_4$ (second row) with 80\% nitrogen in different ambient fields at the end of the
simulations.} \label{dens_4.fig}
\end {figure}

%%%%%%%%%%%%%%%%%%%%%%%%%%%%%%%%%%%%%%%%%%%%%%%%%%%%%%%%%%%%%%%%%%%%%%

%%%%%%%%%%%%%%%%%%%%%%%FIG. 4%%%%%%%%%%%%%%%%%%%%%%%%%%%%%%%%%%%%%%%%%
\begin {figure}
\begin {center}
\begin {tabular}{cc}
\includegraphics [scale=0.56] {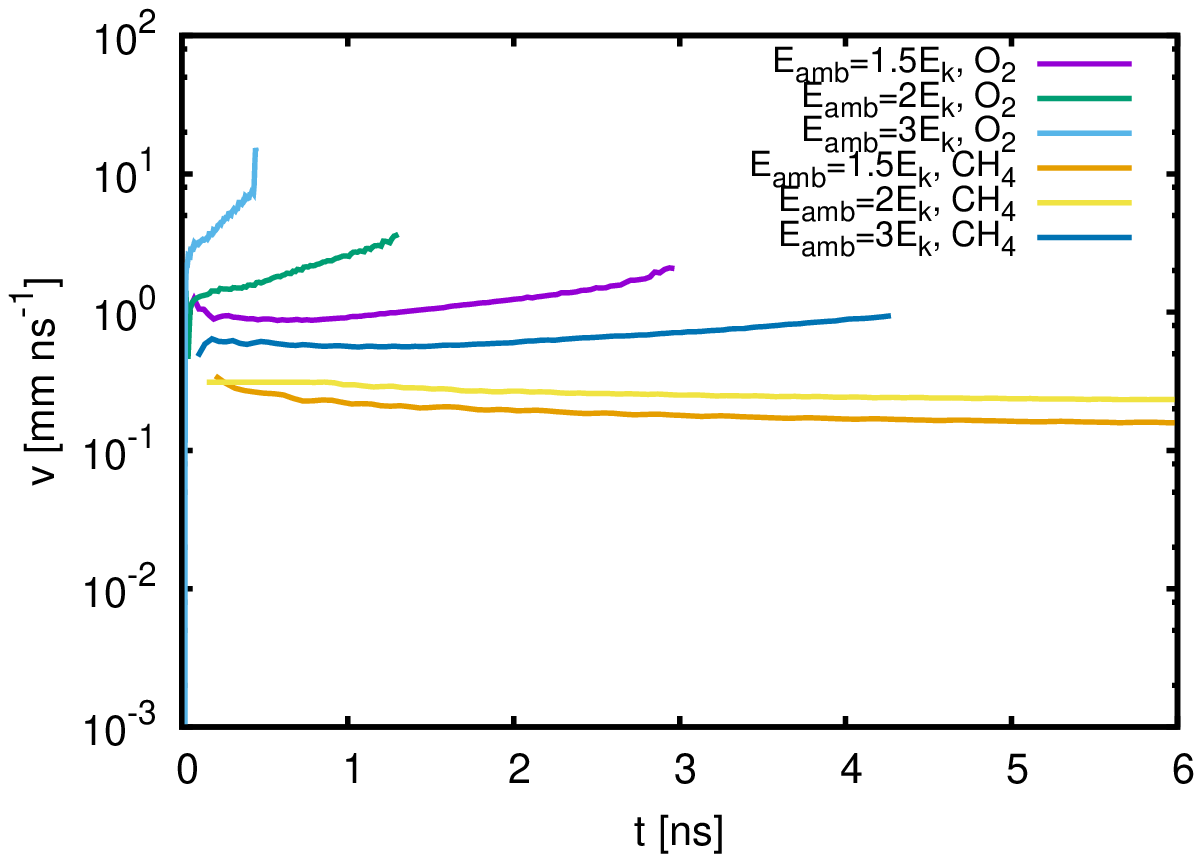} &
\includegraphics [scale=0.56] {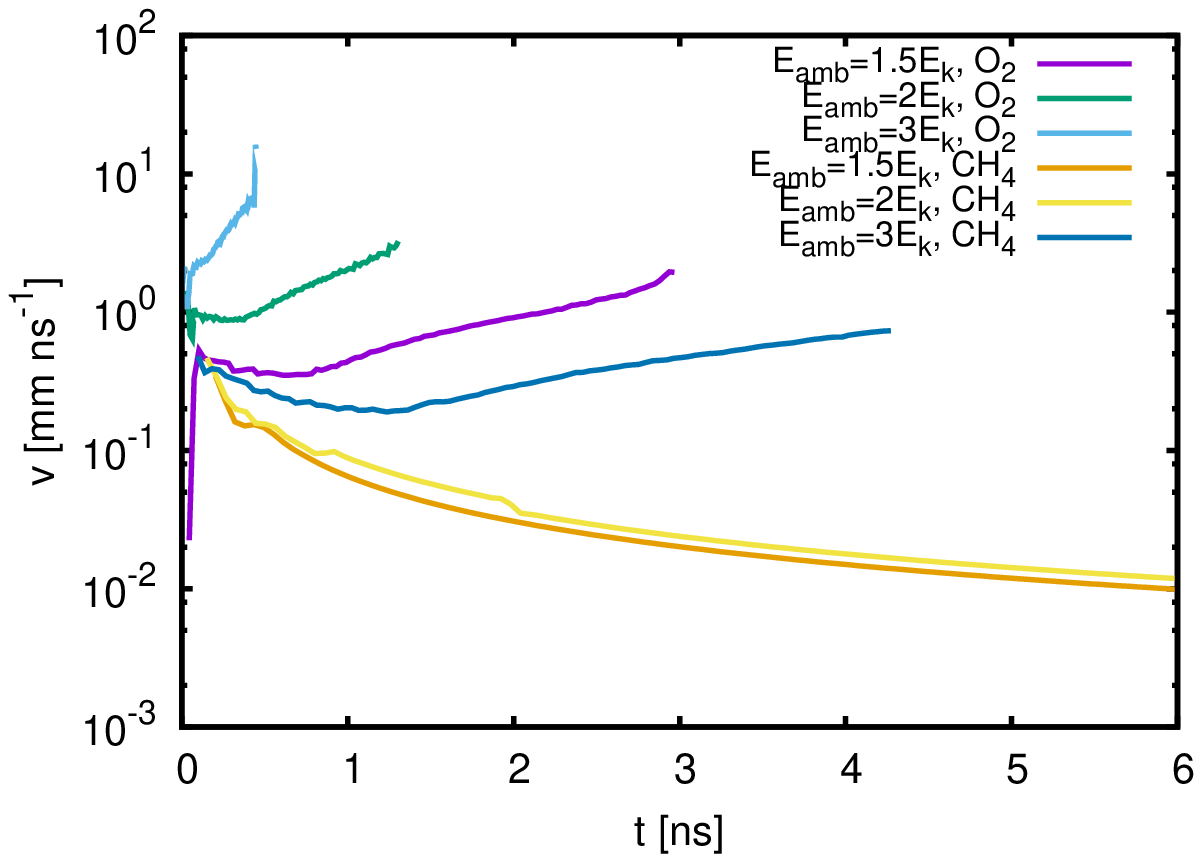} \\
a) $\kappa=0.8$, negative & b) $\kappa=0.8$, positive
\end {tabular}
\end {center}
\caption {The velocity of the positive (first column) and negative
front (second column) as a function of time for gas mixtures with 80\% nitrogen.} \label{veloc.fig}
\end {figure}
%%%%%%%%%%%%%%%%%%%%%%%%%%%%%%%%%%%%%%%%%%%%%%%%%%%%%%%%%%%%%%%%%%%%%%

%%%%%%%%%%%%%%%%%%%%%%%FIG. 5%%%%%%%%%%%%%%%%%%%%%%%%%%%%%%%%%%%%%%%%%
\begin {figure}
\begin {center}
\begin {tabular}{cc}
\includegraphics [scale=0.30] {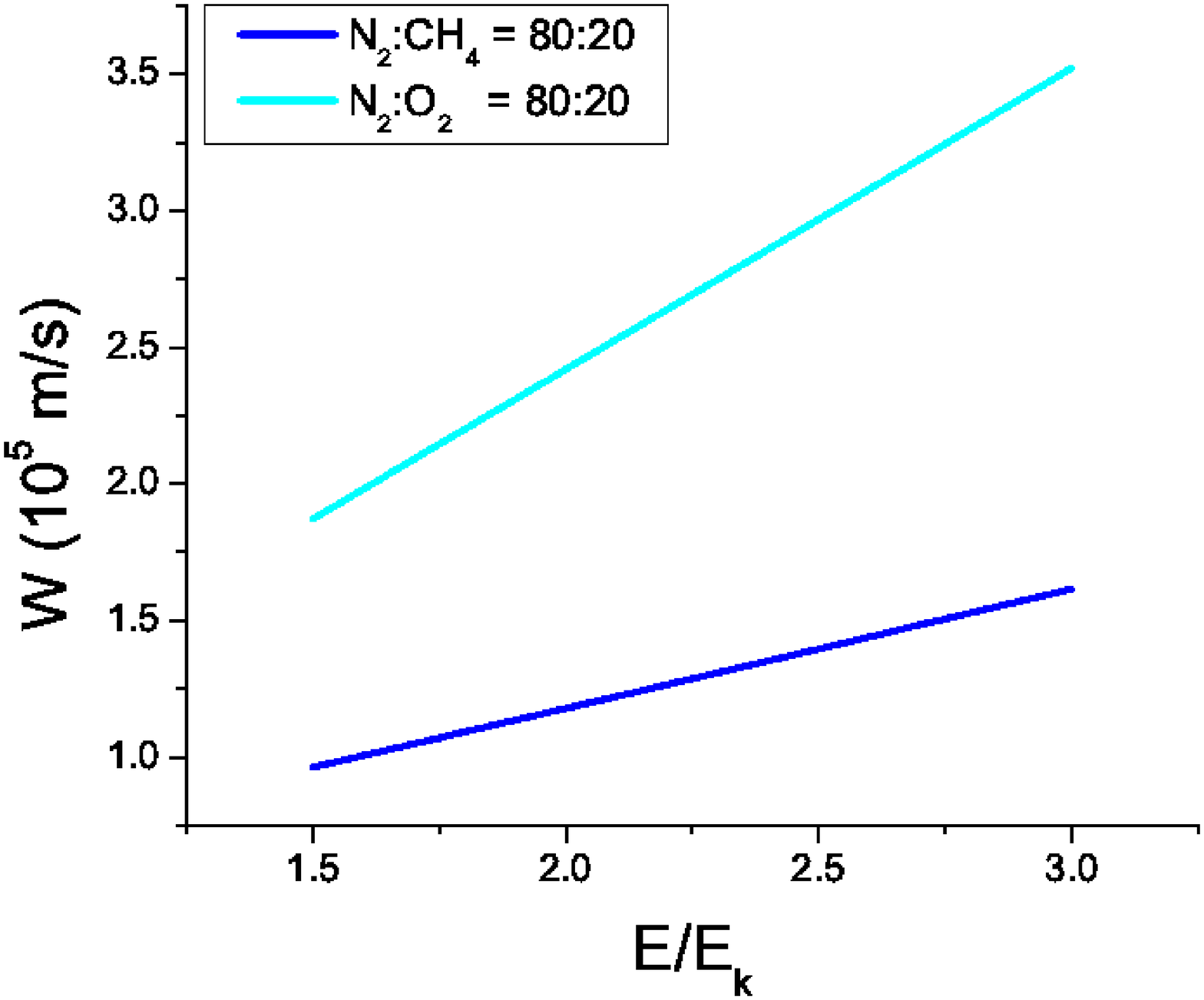} &
\includegraphics [scale=0.30] {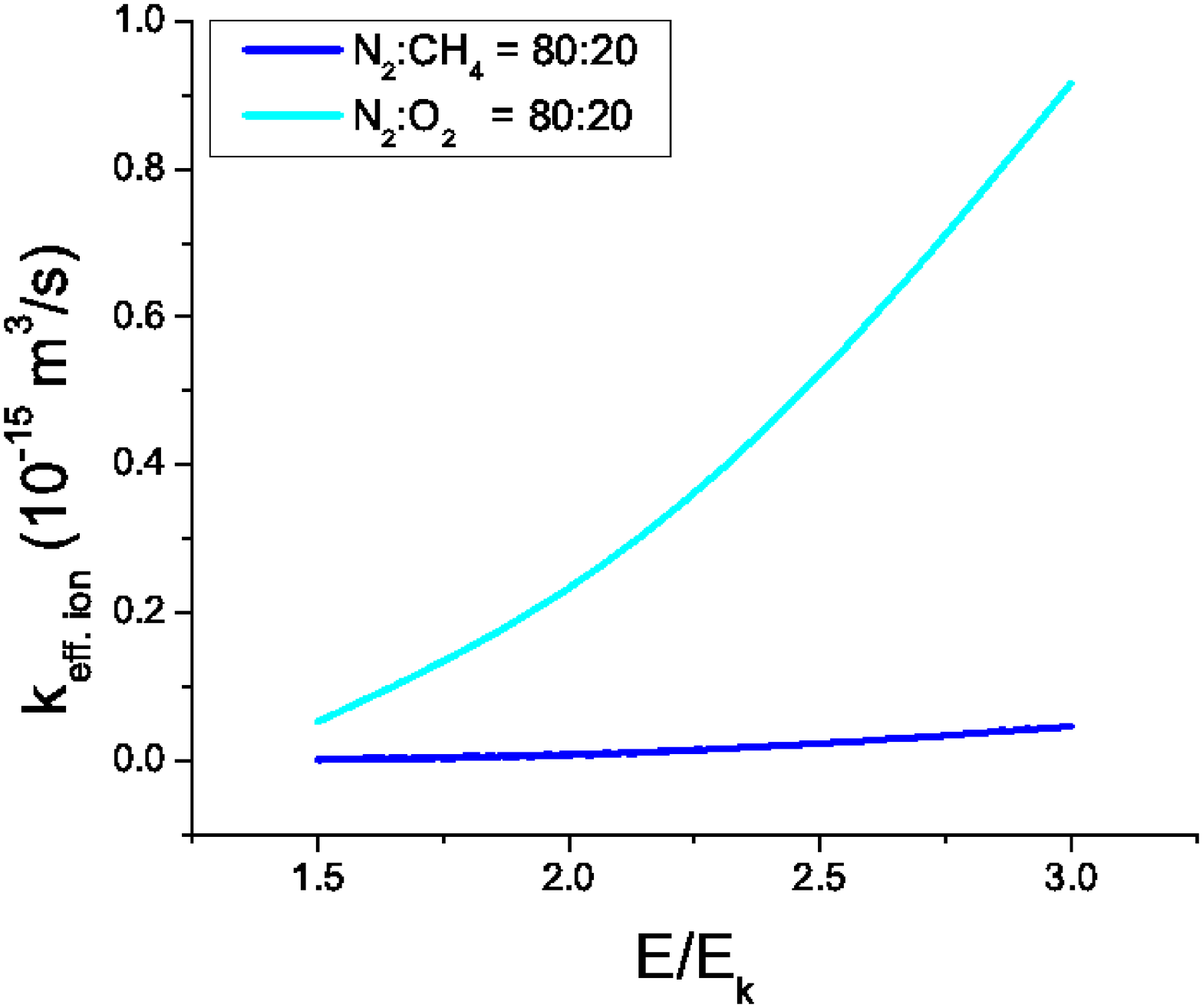} \\
a) & b) \\
\includegraphics [scale=0.56] {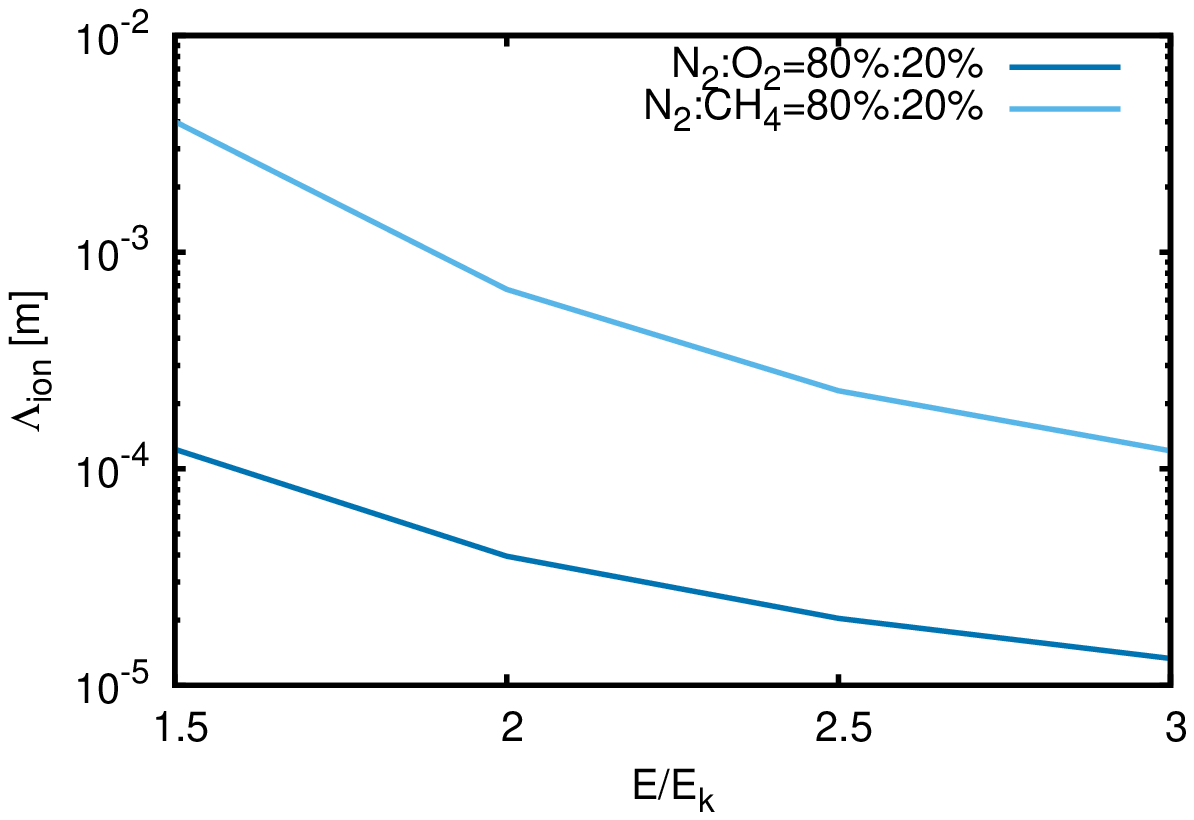}
c)
\end {tabular}
\end {center}
\caption {The drift velocity (a), the effective ionization
coefficient (b) and the ionization length $\Lambda_{ion}$ (c) as a function of the ratio of the applied electric field and
breakdown field.} \label{coef.fig}
\end {figure}
%%%%%%%%%%%%%%%%%%%%%%%%%%%%%%%%%%%%%%%%%%%%%%%%%%%%%%%%%%%%%%%%%%%%%%

%%%%%%%%%%%%%%%%%%%%%%%FIG. 6%%%%%%%%%%%%%%%%%%%%%%%%%%%%%%%%%%%%%%%%%
\begin {figure}
\begin {center}
\begin {tabular}{cc}
\includegraphics[scale=0.4] {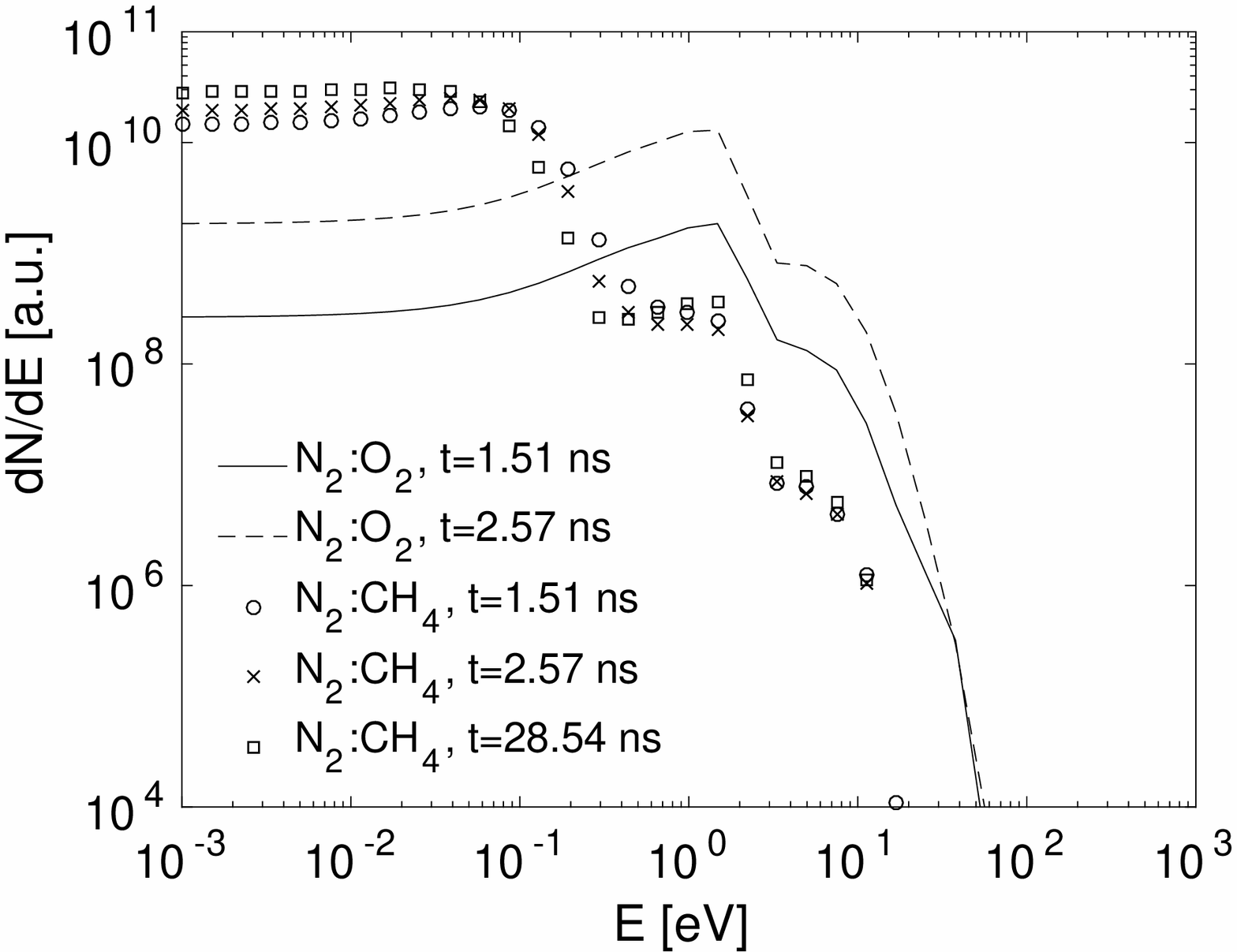} &
\includegraphics[scale=0.4] {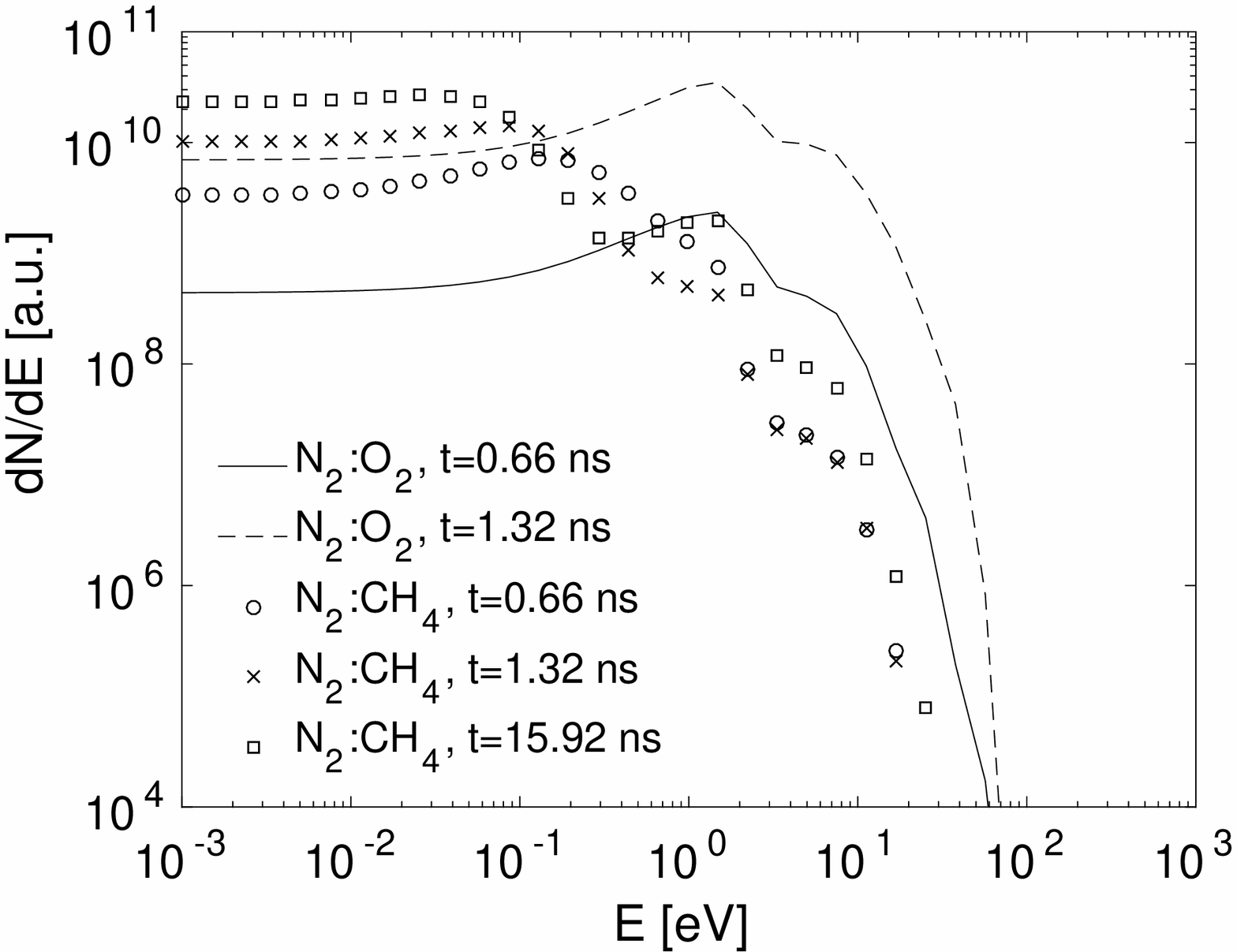} \\
a) $E_{amb}=1.5E_k$ & b) $E_{amb}=2E_k$\\
\vspace{0.2cm}\\
\includegraphics[scale=0.4] {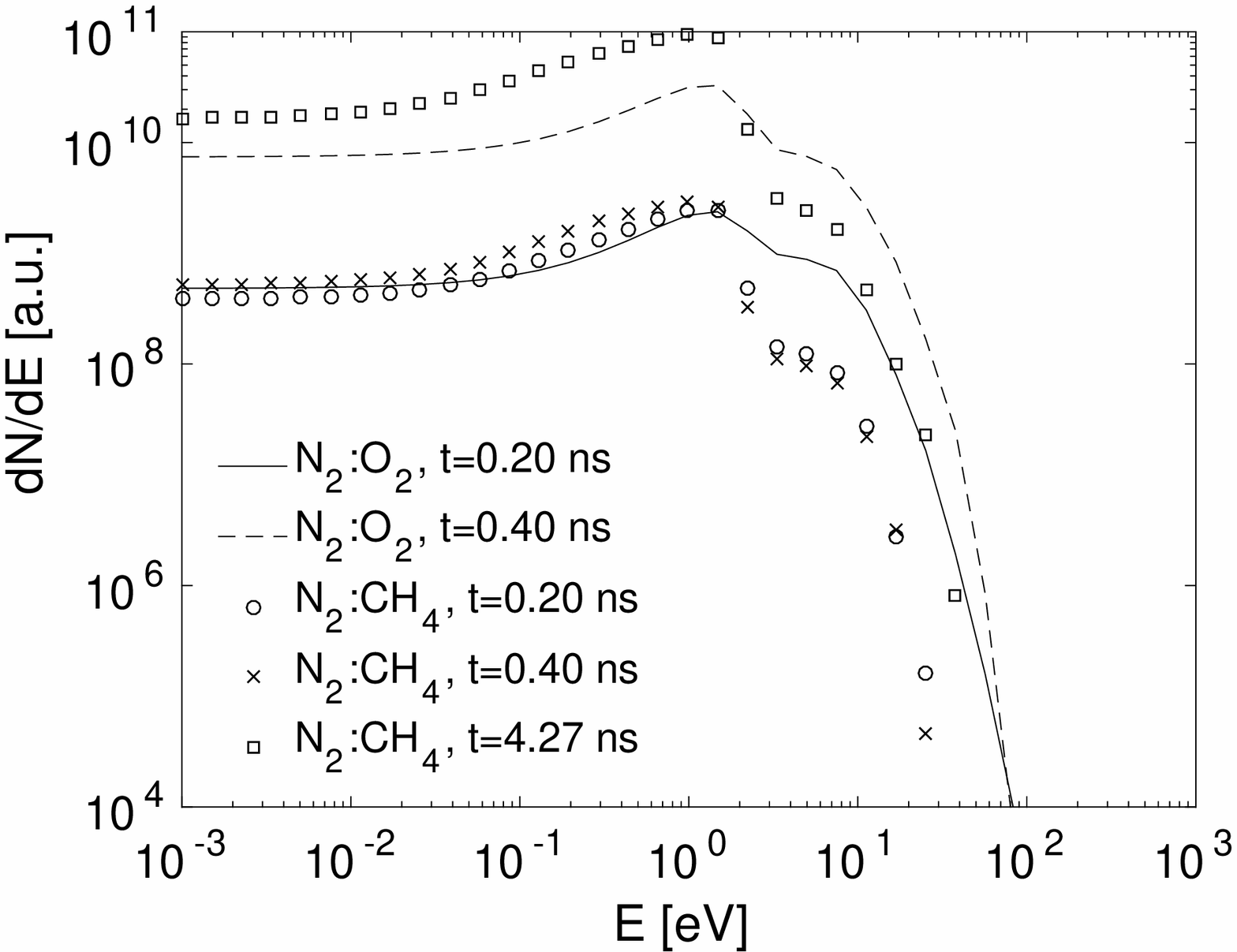} &
\includegraphics[scale=0.4] {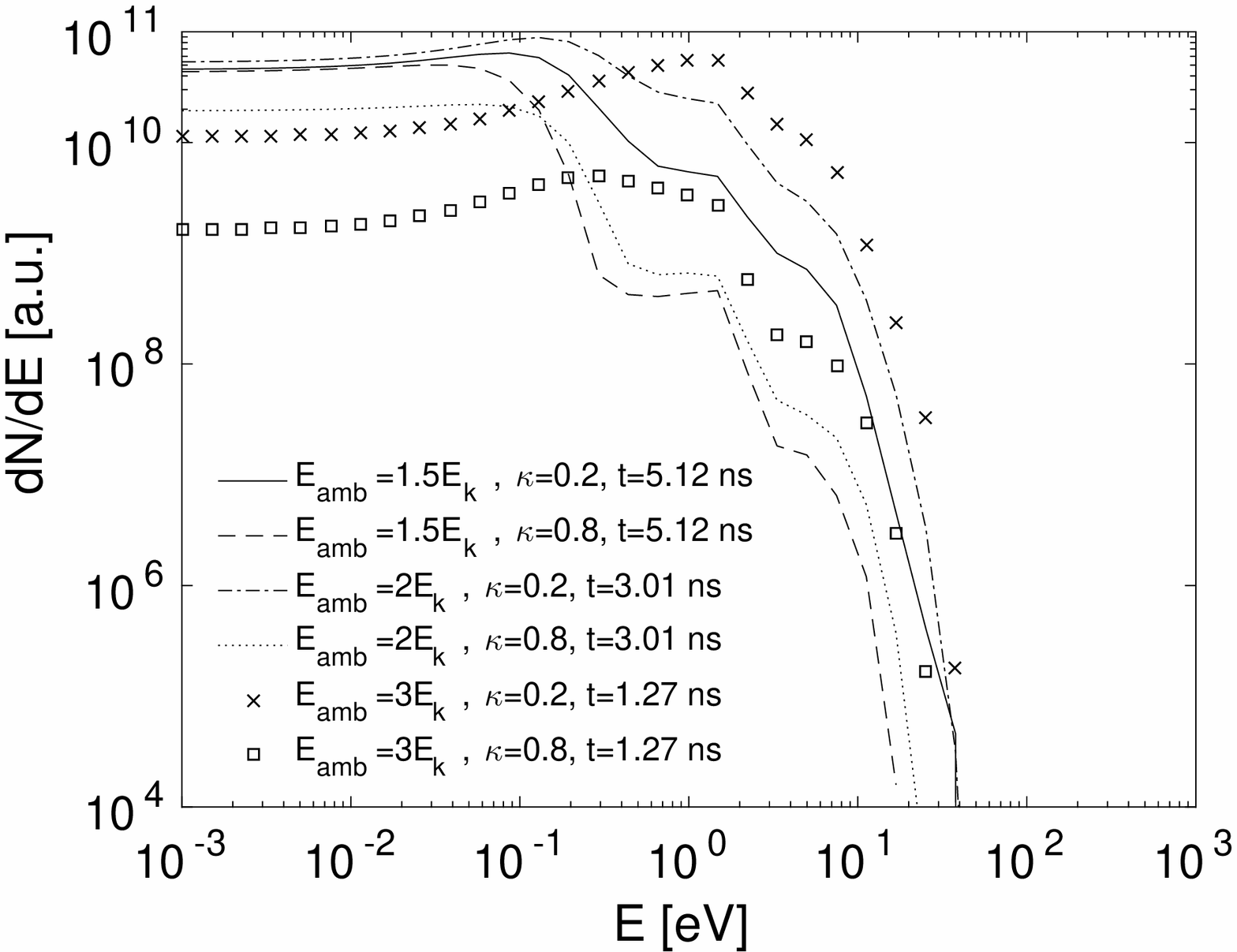} \\
c) $E_{amb}=3E_k$ & d) $\kappa=0.2,0.8$
\end {tabular}
\end {center}
\caption {The electron energy distribution in an ambient field of a)
$1.5E_k$, b) $2E_k$ and c) $3E_k$ for $\kappa=0.8$ after the same time steps as in Fig.
\ref{dens_1.fig}. d) The energy distribution for $\kappa=0.2$ and for
$\kappa=0.8$ for the same fields and time steps as in Fig. \ref{dens_2.fig}.} \label{energy_1.fig}
\end {figure}
%%%%%%%%%%%%%%%%%%%%%%%%%%%%%%%%%%%%%%%%%%%%%%%%%%%%%%%%%%%%%%%%%%%%%%

%%%%%%%%%%%%%%%%%%%%%%%FIG. 7%%%%%%%%%%%%%%%%%%%%%%%%%%%%%%%%%%%%%%%%%
\begin {figure}
\begin {center}
\begin {tabular}{ccc}
\includegraphics [scale=0.3] {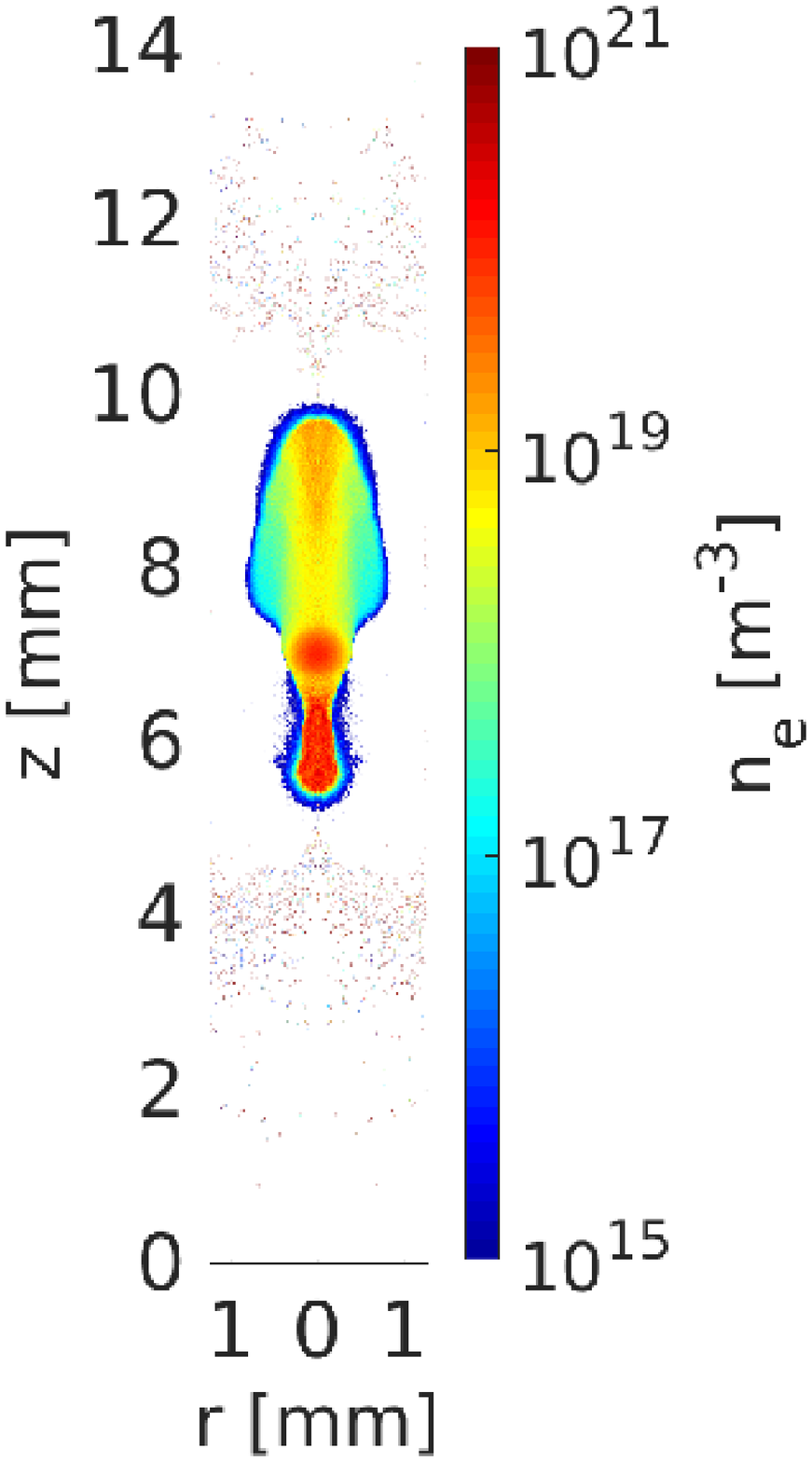} &
\includegraphics [scale=0.3] {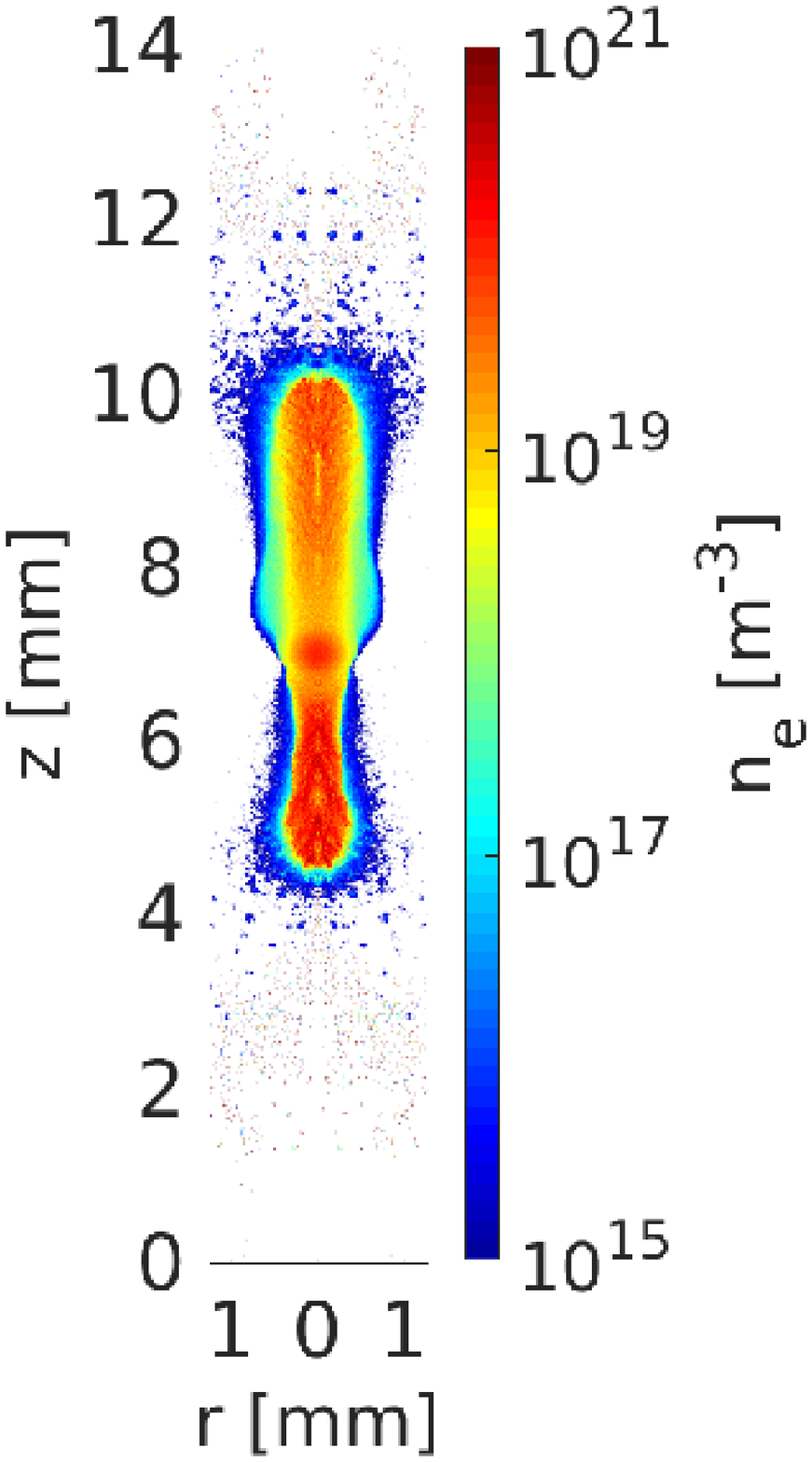} &
\includegraphics [scale=0.3] {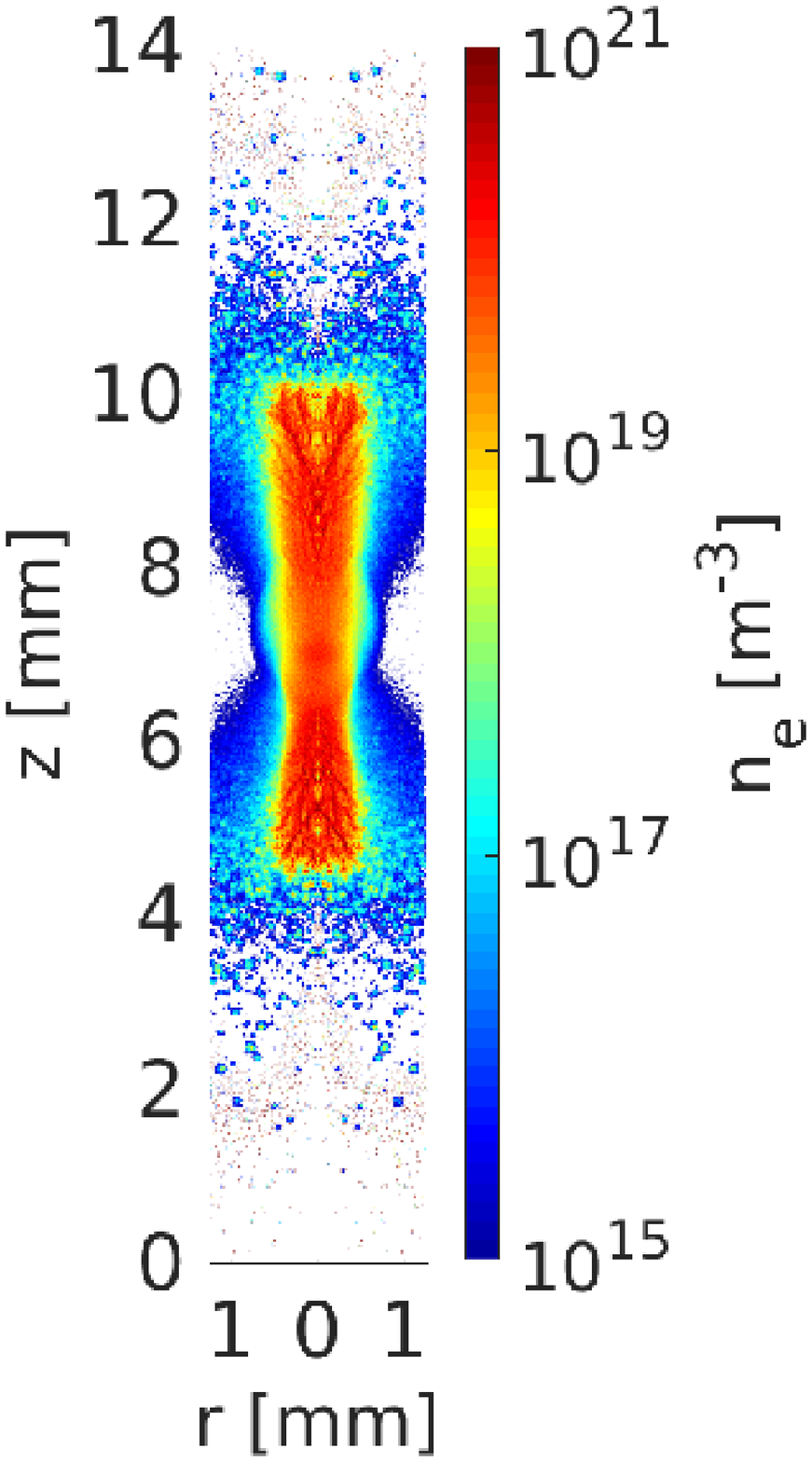} \\
\includegraphics [scale=0.3] {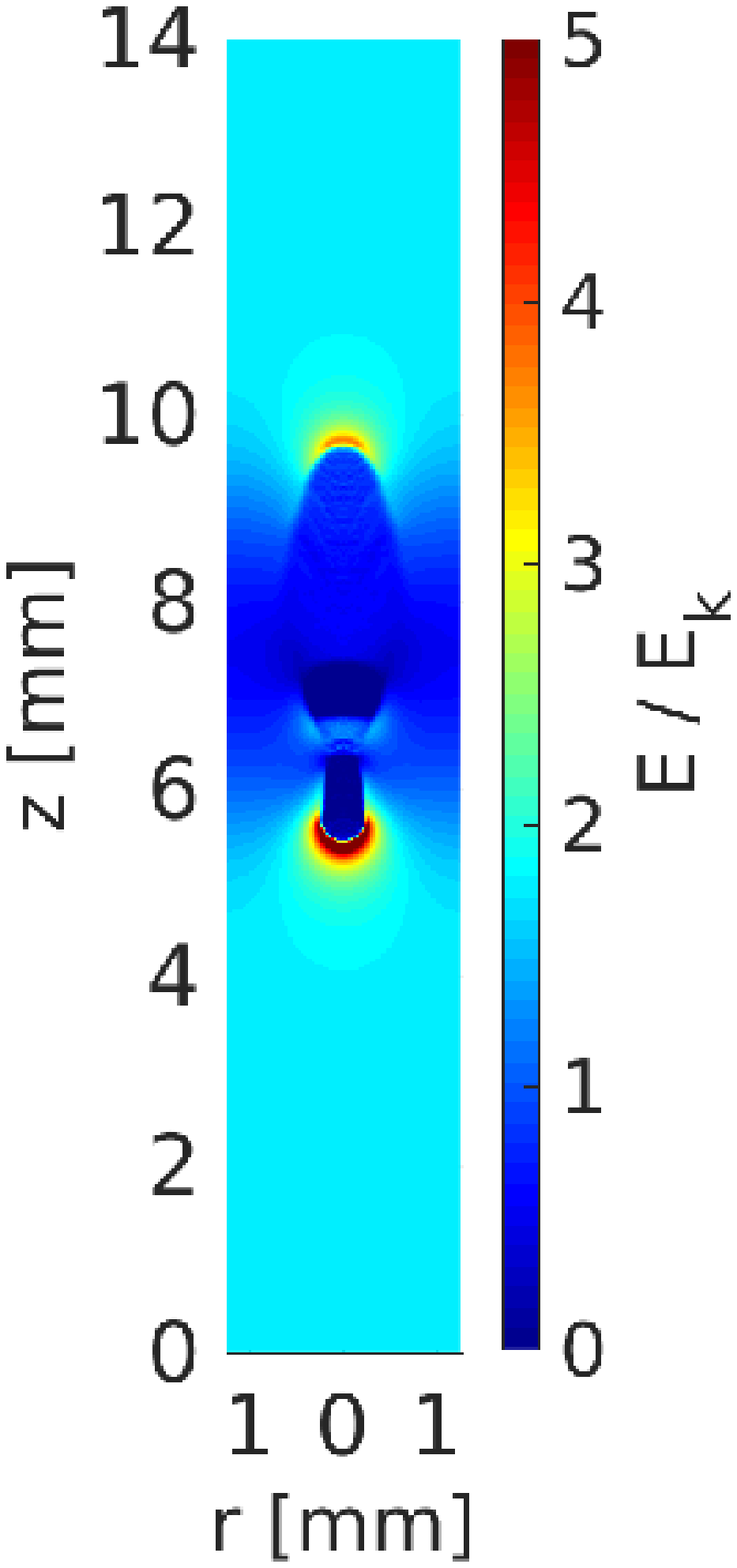} &
\includegraphics [scale=0.3] {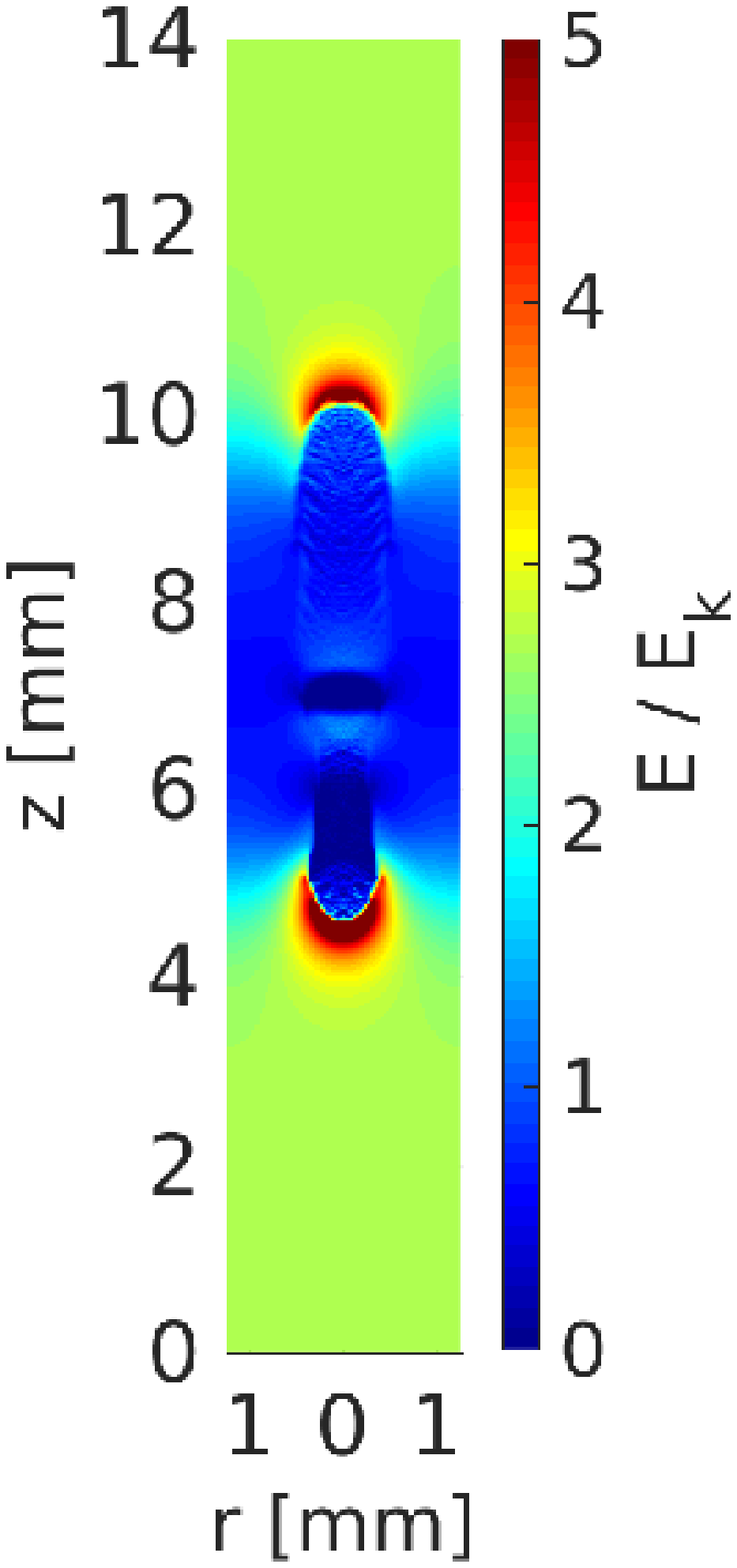} &
\includegraphics [scale=0.3] {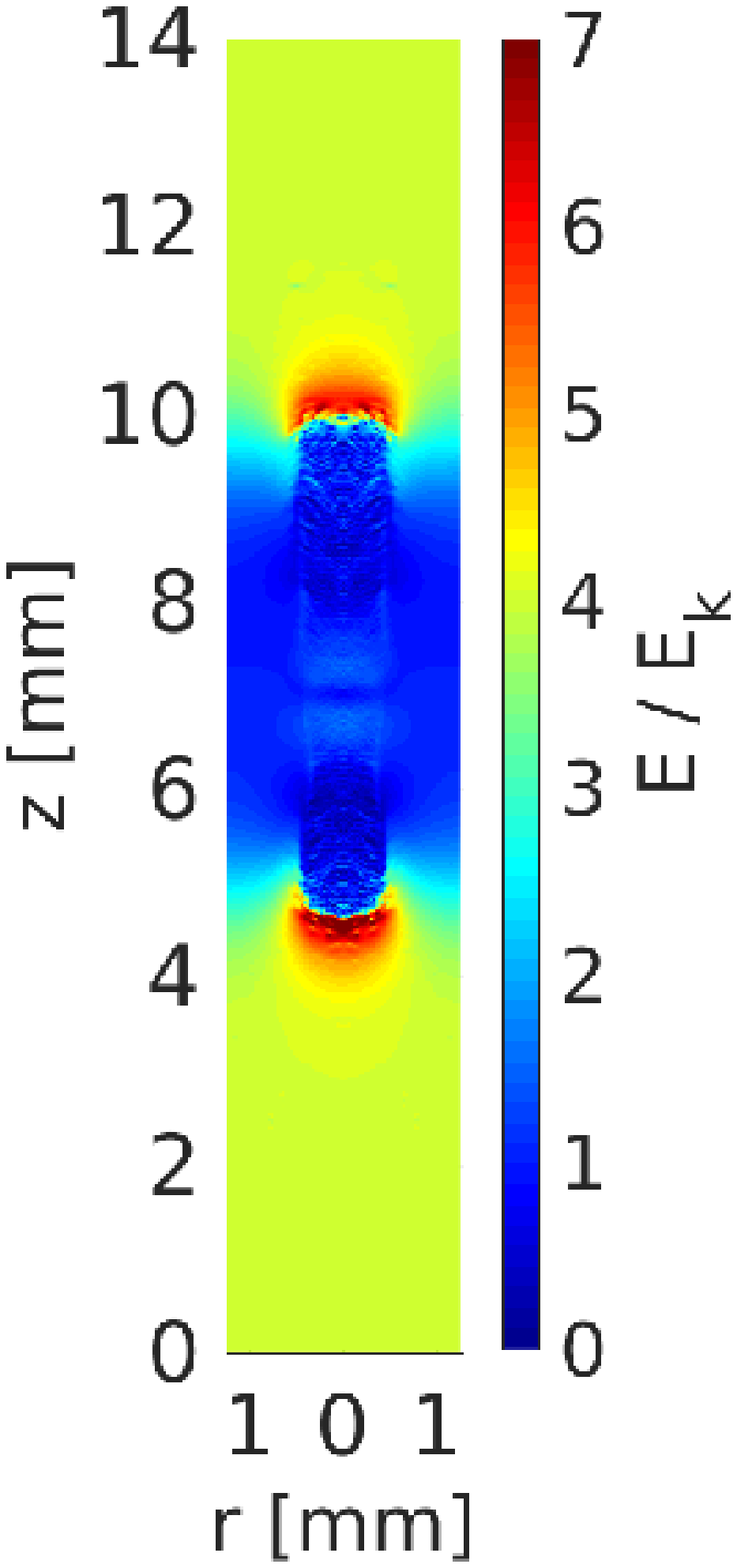} \\
$E=1.5E_k, t=5.12$ ns & $E=2E_k, t=3.01$ ns & $E=3E_k, t=1.27$ ns
\end {tabular}
\end {center}
\caption {The electron density (first row) and the electric field (second
row) in N$_2$:CH$_4$=(20\%):(80\%) in different electric fields after different
time steps.} \label {dens_2.fig}
\end {figure}

%%%%%%%%%%%%%%%%%%%%%%%%%%%%%%%%%%%%%%%%%%%%%%%%%%%%%%%%%%%%%%%%%%%%%%

%%%%%%%%%%%%%%%%%%%%%%%FIG. 8%%%%%%%%%%%%%%%%%%%%%%%%%%%%%%%%%%%%%%%%%
\begin {figure}
\begin {center}
\begin {tabular}{cc}
\includegraphics [scale=0.56] {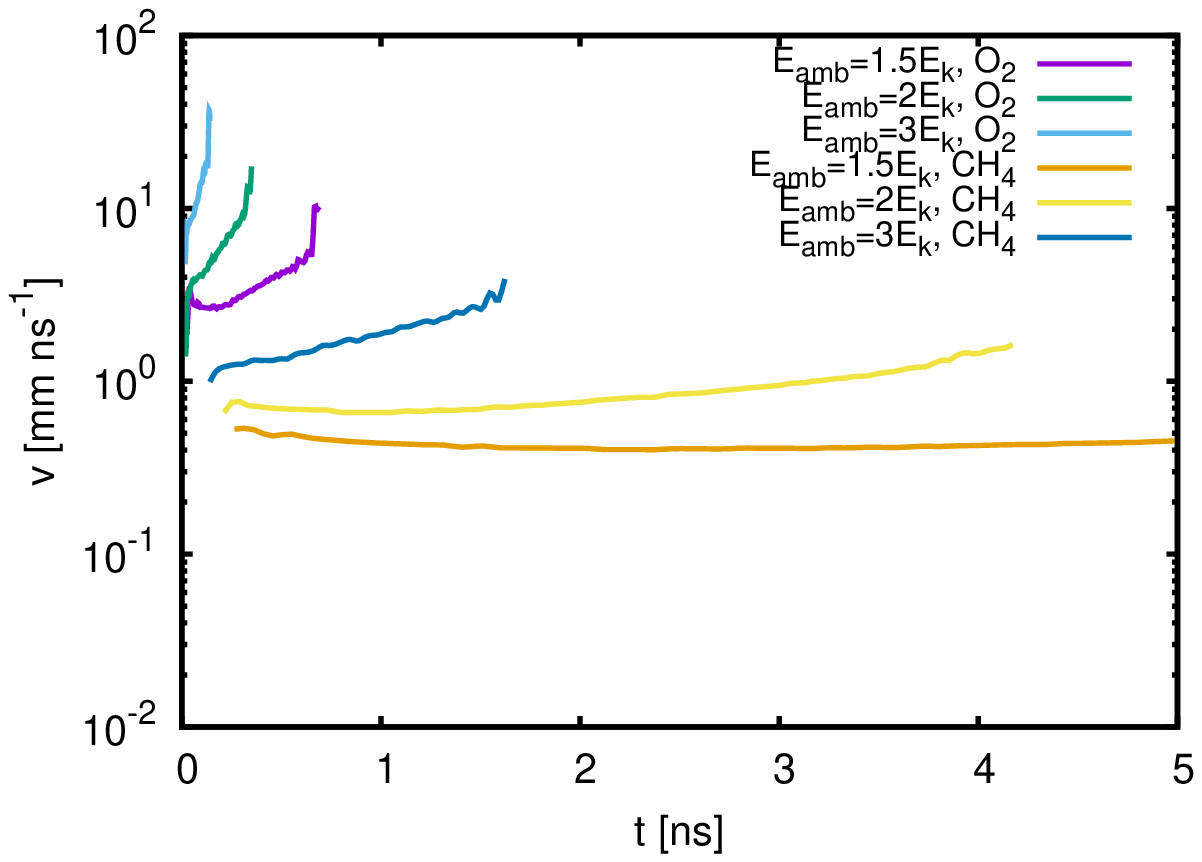} &
\includegraphics [scale=0.56] {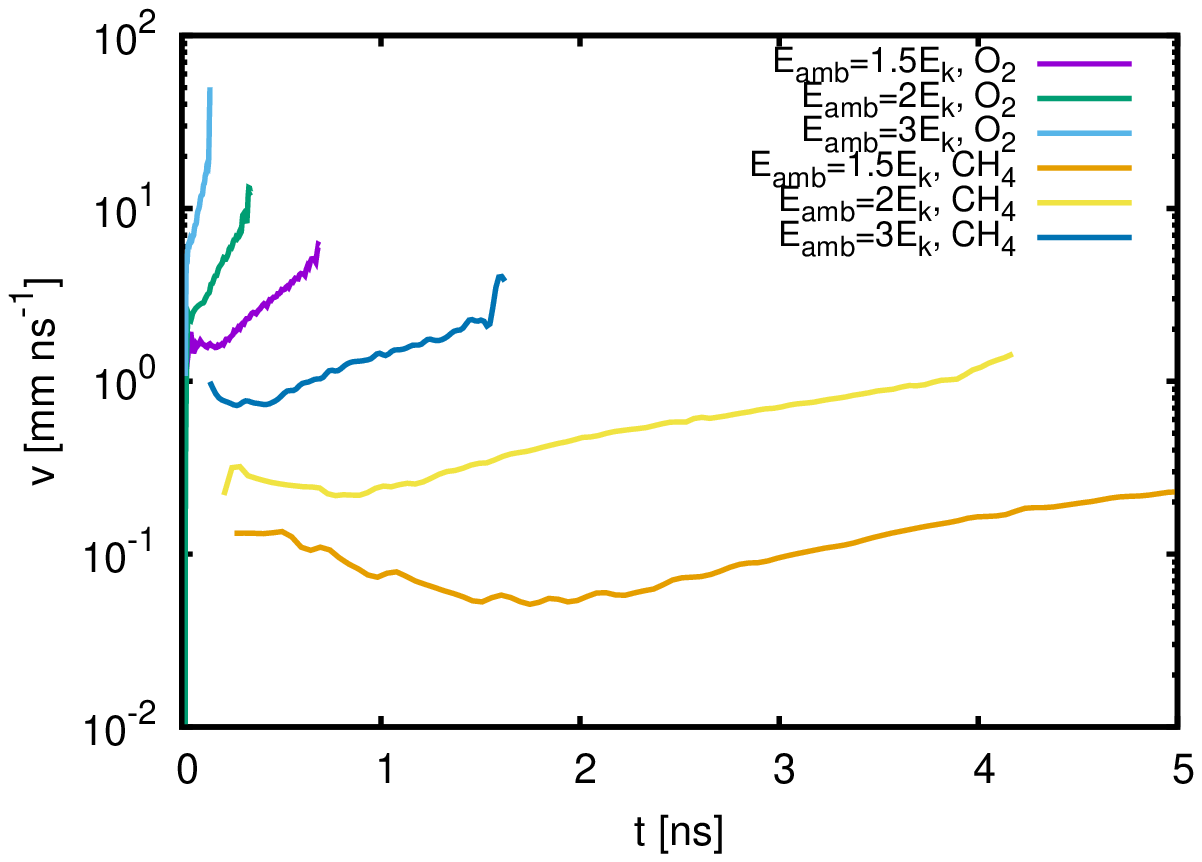} \\
a) $\kappa=0.2$, negative & b) $\kappa=0.2$, positive \\
\includegraphics [scale=0.56] {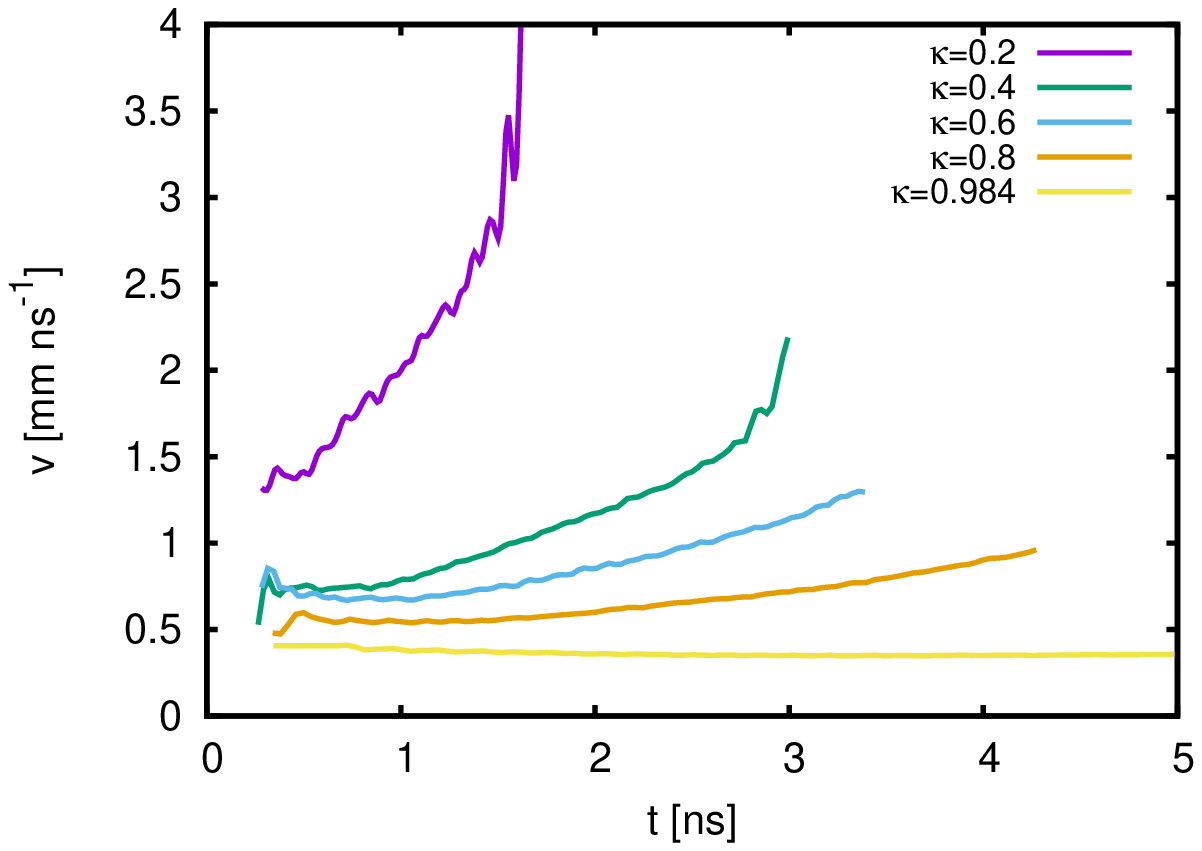} &
\includegraphics [scale=0.56] {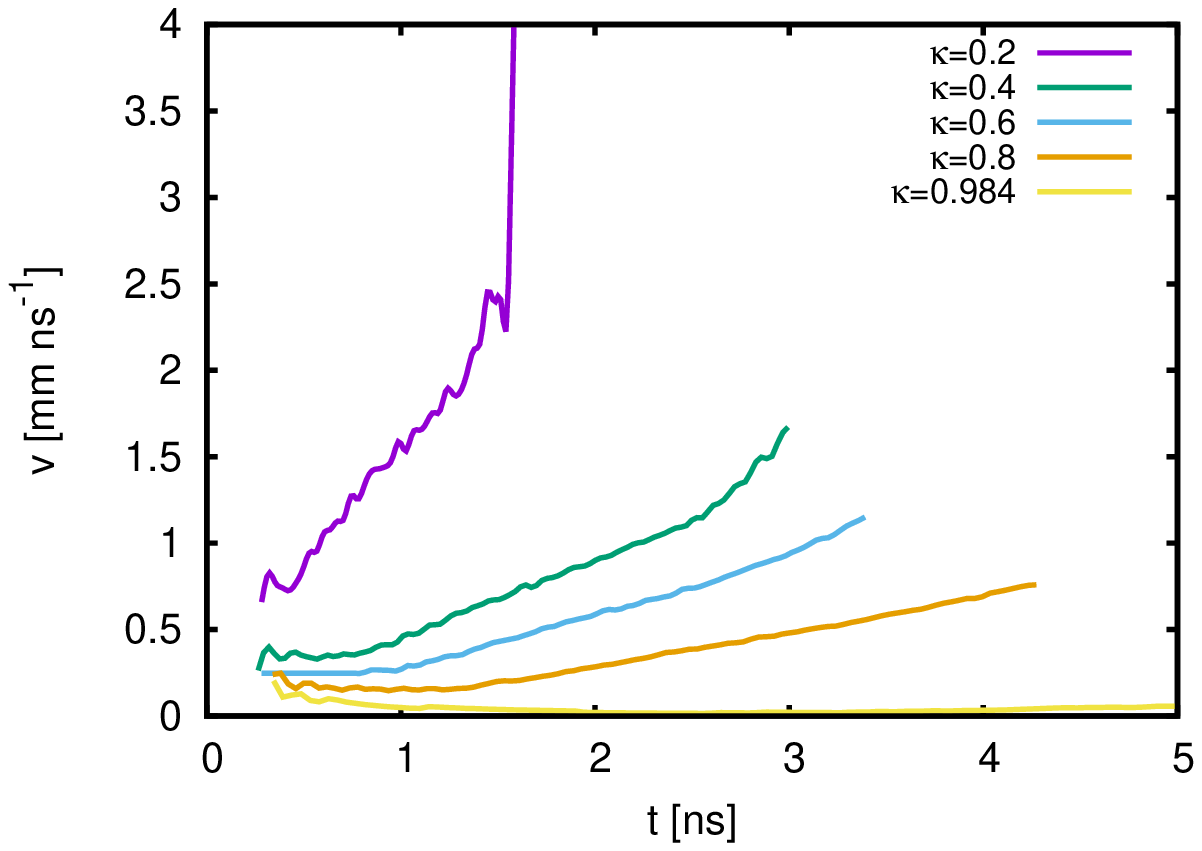} \\
c) $E_{amb}=3E_k$, negative & d) $E_{amb}=3E_k$, positive 
\end {tabular}
\end {center}
\caption {a,b) The velocity of the negative (a) and positive
front (b) as a function of time for gas mixtures with 20\% nitrogen.
c,d) The velocity of the negative and of the positive front in N$_2$:CH$_4$
in an ambient field of $3E_k$ for different percentages of nitrogen.}
\label{veloc_2.fig}
\end {figure}
%%%%%%%%%%%%%%%%%%%%%%%%%%%%%%%%%%%%%%%%%%%%%%%%%%%%%%%%%%%%%%%%%%%%%%

Under the influence of the ambient field the initial electron-ion patch first
deve\-lops into an electron avalanche and eventually into a bidirectional
streamer. The ambient electric field points downwards such that the negative front moves upwards whilst the positive front moves downwards. The attached video files show the temporal evolution of the electron density
and the electric field for the gas mixtures with 20\% and 1.6\% methane.

Figure \ref{dens_1.fig} shows the electron density and the electric field
(third column) for gas mixtures with 80\% nitrogen in
ambient fields of $1.5E_k$ (first row), $2E_k$ (second row), $3E_k$ (third
row), where $E_k$ is $\approx 3.5$ MV m$^{-1}$ in mixtures with oxygen
and $\approx 1.6$ MV m$^{-1}$ in mixtures with methane, after different time steps. In columns one to three, the left halves of each panel show the electron density or
electric field in mixtures with 20\% oxygen, i.e. the concentration on Earth whereas the right
halves show the density and field in mixtures with 20\% methane which we refer to as Earth-like Titan.
Figure \ref{dens_4.fig} shows the electron density
on Earth and in Earth-like Titan mixtures at the end of
our simulations.

In our simulations, the electron-ion patch on Earth develops into a
bidirectional streamer for any field above $1.5E_k$. The electric field shows
the typical streamer-like pattern with an enhanced field at the tips and a shielded
field inside the streamer body. Exchanging oxygen with methane dramatically
changes the picture; the process of forming electron avalanches or
eventually streamers becomes delayed significantly irrespective of the applied
field. Columns one and two show that
there is hardly any development whilst in mixtures with oxygen, streamers have already formed after the same time steps. However, the electric field shows
the same pattern as in N$_2$:O$_2$ with an enhanced
field at the tips of the electron patch and a vanishing shield in the patch's
interior.

At the end of the simulations we observe three different scenarios in mixtures with CH$_4$ depending
on the ambient electric field. In a field of
$1.5E_k$, the field at the positive front increases; however, the positive
front does not move, hence there is no development of a positive streamer
front. On the negative side, an electron avalanche, but no streamer front forms, without any
significant field increase at the top. At the end of the simulation, the
field at the tip is $1.8E_k$ and the electron density is $\lesssim 10^{17}$ m$^{-3}$.
In an ambient field of $2E_k$, the situation does not change
significantly at the positive front. The field increases, but the front does
not develop. On the negative front, however, we observe the formation of a
streamer-like channel. During the simulation, the field at the tip varies between
approximately $3E_k$ and $6E_k$ whereas the field is shielded in its interior. In contrast
to the density in $1.5E_k$ (d), the electron number slightly multiplies and reaches a density of
$\approx 10^{19}$ m$^{-3}$ after 15.92 ns. In $E_{amb}=3E_k$, there is the distinct formation
of a positive and of a negative streamer front similar to as in air. Since the
breakdown fields in N$_2$:O$_2$ and in N$_2$:CH$_4$ differ by a factor of
$\approx 2$, the absolute value of $3E_k$ in N$_2$:CH$_4$ corresponds to the value of $1.5E_k$ in air. Hence,
the evolution of the electron density and of the electric field is
comparable although delayed. 

Figure \ref{veloc.fig} shows the front velocities in the early stages
of the evolution of electron avalanches and streamers for
$\kappa=0.8$ for O$_2$ and CH$_4$ and for all considered electric fields. Depending on the electric
field, the streamer velocity in N$_2$:O$_2$ lies between $10^0$ and $10^1$ mm ns$^{-1}$ and is
similar for the positive and the negative streamer front. For $E_{amb}=1.5E_k$ and
$E_{amb}=2E_k$ in N$_2$:CH$_4$ the velocity of the positive front lies between
$10^{-2}$ and $10^{-1}$ mm ns$^{-1}$ and is decreasing with time which is in agreement with Figures
\ref{dens_1.fig} d) and h) with no distinct fronts for these fields, similarly for negative fronts. For $3E_k$,
Fig. \ref{dens_1.fig} l) shows the development of a positive and a negative
streamer front. The slope of the velocity is comparable to the increase
of streamer velocities in air in a field of $1.5E_k$; however, in air, positive fronts and
negative fronts are faster than in N$_2$:CH$_4$. The delayed motion of
fronts in mixtures with methane is related to the drift of electrons.
Fig. \ref{coef.fig} a) shows the drift velocity of electrons in nitrogen-methane
and nitrogen-oxygen mixtures as a function of the ambient field. It is
higher and increasing more significantly in mixtures with oxygen.
Consequently the motion of fronts in N$_2$:CH$_4$ mixtures is delayed.

The differences in the avalanche-to-streamer transition as well as in the
streamer evolution between Earth and Earth-like Titan for the same $E_{amb}/E_k$
results from different
ionization lengths in N$_2$:O$_2$ and in
N$_2$:CH$_4$ as well as from the reduced probability for photoionization.
Figure \ref{coef.fig} b) shows the effective ionization
coefficient as a function of the electric field. It shows that
ionization in a mixture with 20\% methane is less effective than in a mixture
with 20\% oxygen for all considered field strengths. Beyond, the growth of
the ionization coefficient is much less dominant in N$_2$:CH$_4$ mixtures
than in N$_2$:O$_2$. Panel c) shows the ionization length $\Lambda_{ion}$ as
a function of $E/E_k$. In N$_2$:O$_2$ mixtures the ionization length varies
from approximately 0.1 mm for $1.5E_k$ to 0.01 mm for $3E_k$ whilst it
amounts to 4 mm for $1.5E_k$ in N$_2$:CH$_4$ mixtures with 80\% nitrogen.
Hence, in $1.5E_k$ in mixtures with methane, the ionization length is
comparable to the size of the simulation domain preventing the occurrence of a
significant amount of ionization. Since ionization is less effective in mixtures with
methane, the build-up of space charges and thus the formation of enhanced
electric field tips is delayed which reduces the electron energies and the
further drive of ionization. Note that the ionization coefficient and the
ionization length for $3E_k$
in N$_2$:CH$_4$ is comparable to the ionization coefficient for $1.5E_k$ in
N$_2$:O$_2$. For the same time steps and fields as in
Fig. \ref{dens_1.fig}, Figure \ref{energy_1.fig} a)-c) show the electron energy
distribution. The solid and dashed lines
show the typical streamer-like energy distributions of electrons in air \cite{chanrion_2008} with maximum
energies of $\approx 50$ eV ($1.5E_k$, a), $\approx 65$ eV ($2E_k$,
b) and $\approx 80$ eV ($3E_k$, c). However, for the same $E_{amb}/E_k$
(remember that $E_k$ is smaller in mixtures with methane), the
maximum electron energies in N$_2$:CH$_4$ are $\approx 15$ eV (a), $\approx 20$ eV (b) and
$\approx 30$ eV (c), thus only a little above the ionization energy of
nitrogen (15.6 eV) and of methane (12.6 eV). Hence, the ionization is not effective enough to create high field tips
to accelerate electrons into the energy regime where they can
further create a substantial electron multiplication. Comparing the
electron numbers below 1 eV in N$_2$:CH$_4$ reveals that in an ambient field
of $1.5E_k$ the electron number does not change significantly because of
inefficient ionization whereas the electron number increases significantly in a
field of $3E_k$.

Figure \ref{dens_2.fig} shows the electron density and electric field
in N$_2$:CH$_4$ with 20\% nitrogen. It shows that in all considered cases negative and positive
streamer fronts form. Fig. \ref{energy_1.fig} d) compares the electron energy
distribution for $\kappa=0.2$ and
$\kappa=0.8$ for the same fields and time steps as in Fig \ref{dens_2.fig}. It shows that in all cases, the maximum electron energy is
larger for $\kappa=0.2$ than for $\kappa=0.8$ which results from the dependence of the friction force on the percentage of nitrogen. Figure
\ref{cross.fig} b) shows that the friction force in N$_2$:CH$_4$ has a resonance at
approximately 2 eV enhanced for $\kappa=0.8$, reducing the electron energies.
Consequently, since the electron energies are larger for
$\kappa=0.2$, the ionization of the ambient gas as well as photoionization are more probable and the
avalanche-to-streamer transition is facilitated.

For $\kappa=0.2$, Figure \ref{veloc_2.fig} a) and b) show the streamer velocities in N$_2$:O$_2$ and N$_2$:CH$_4$ for all
considered electric fields as a function of time. Although there is
an avalanche-to-streamer transition for all field strengths, streamers move
faster in oxygen than in methane. As for $\kappa=0.8$, the streamer
velocities equal $\approx 10^0-10^1$ mm ns$^{-1}$ in oxygen whilst they lie
between $10^{-1}$ and approximately $10^0$ mm ns$^{-1}$ in the nitrogen-methane mixture.
Panels c) and d) additionally show the front velocities in N$_2$:CH$_4$ in
an ambient field of $3E_k$ for different percentages of nitrogen. They
illustrate that both positive and negative fronts move faster in mixtures
with a low percentage of nitrogen or equivalently with a significant
contribution of methane. Since the breakdown electric field does not vary much as a function of $\kappa$, we explain this
with the friction force and
its dependency on $\kappa$. As we have discussed in section
\ref{cross.sec}, the friction force
has a resonance at an electron energy of approx. 2 eV which is distinct for
large percentages of nitrogen. Subsequently, electron energies are lower and
the fronts move slower in mixtures with high
percentages of nitrogen. Equivalently fronts move substantially faster in
mixtures with a significant concentration of methane.

\subsection {The inception of streamers on Titan} \label {titan.sec}

%%%%%%%%%%%%%%%%%%%%%%%FIG. 9%%%%%%%%%%%%%%%%%%%%%%%%%%%%%%%%%%%%%%%%%
\begin {figure}
\begin {center}
\begin {tabular}{cccc}
\includegraphics [scale=0.3] {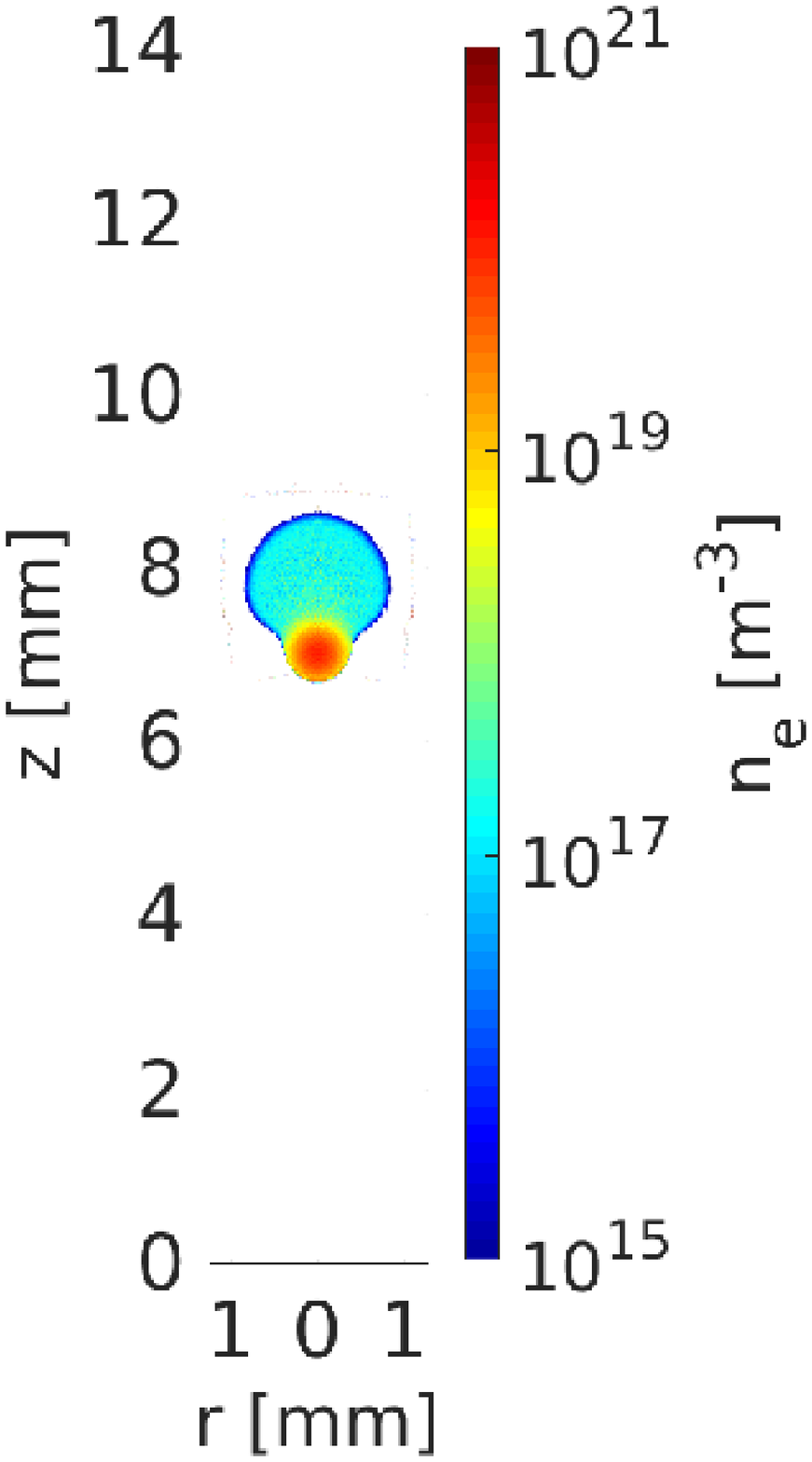} &
\includegraphics [scale=0.3] {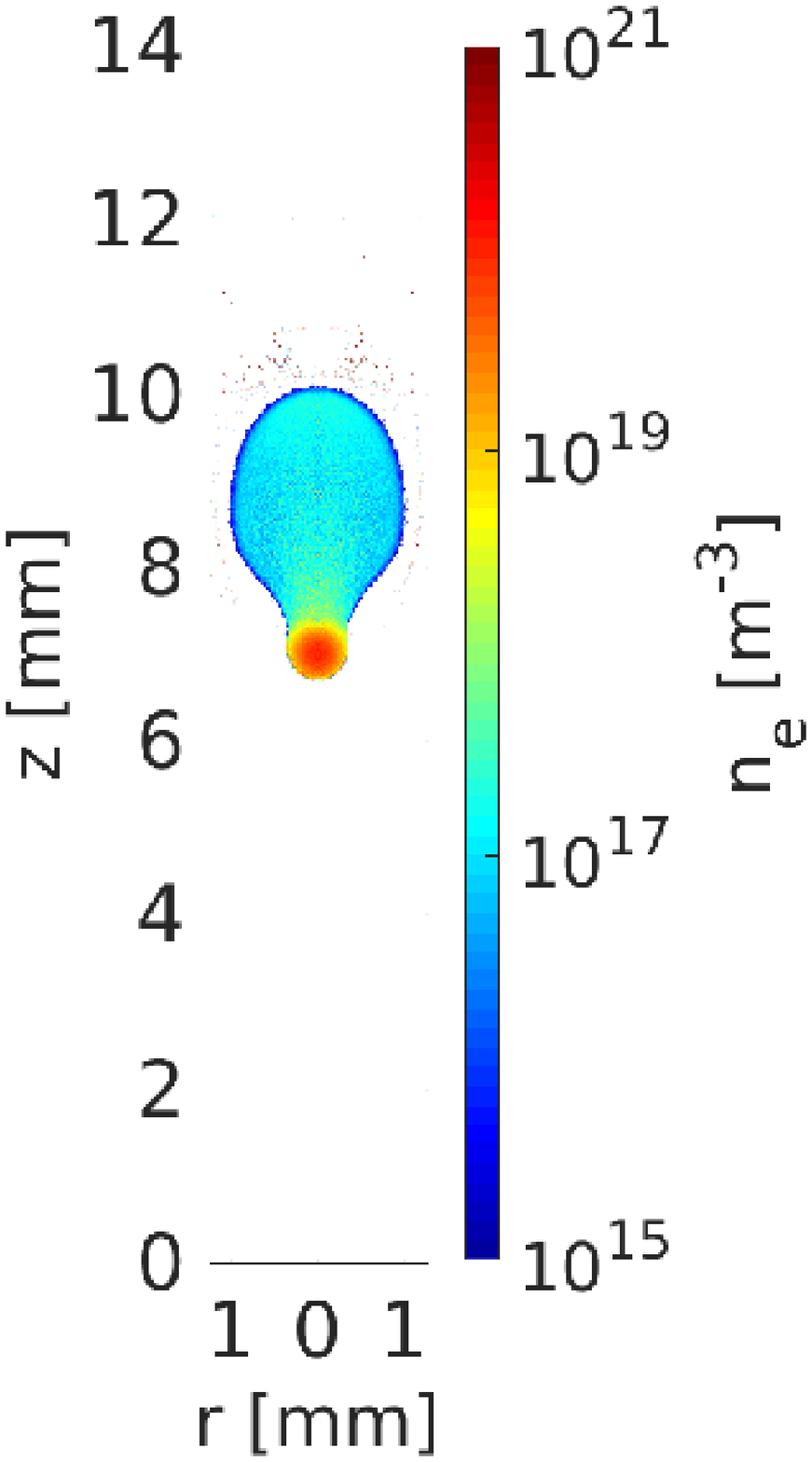} &
\includegraphics [scale=0.3] {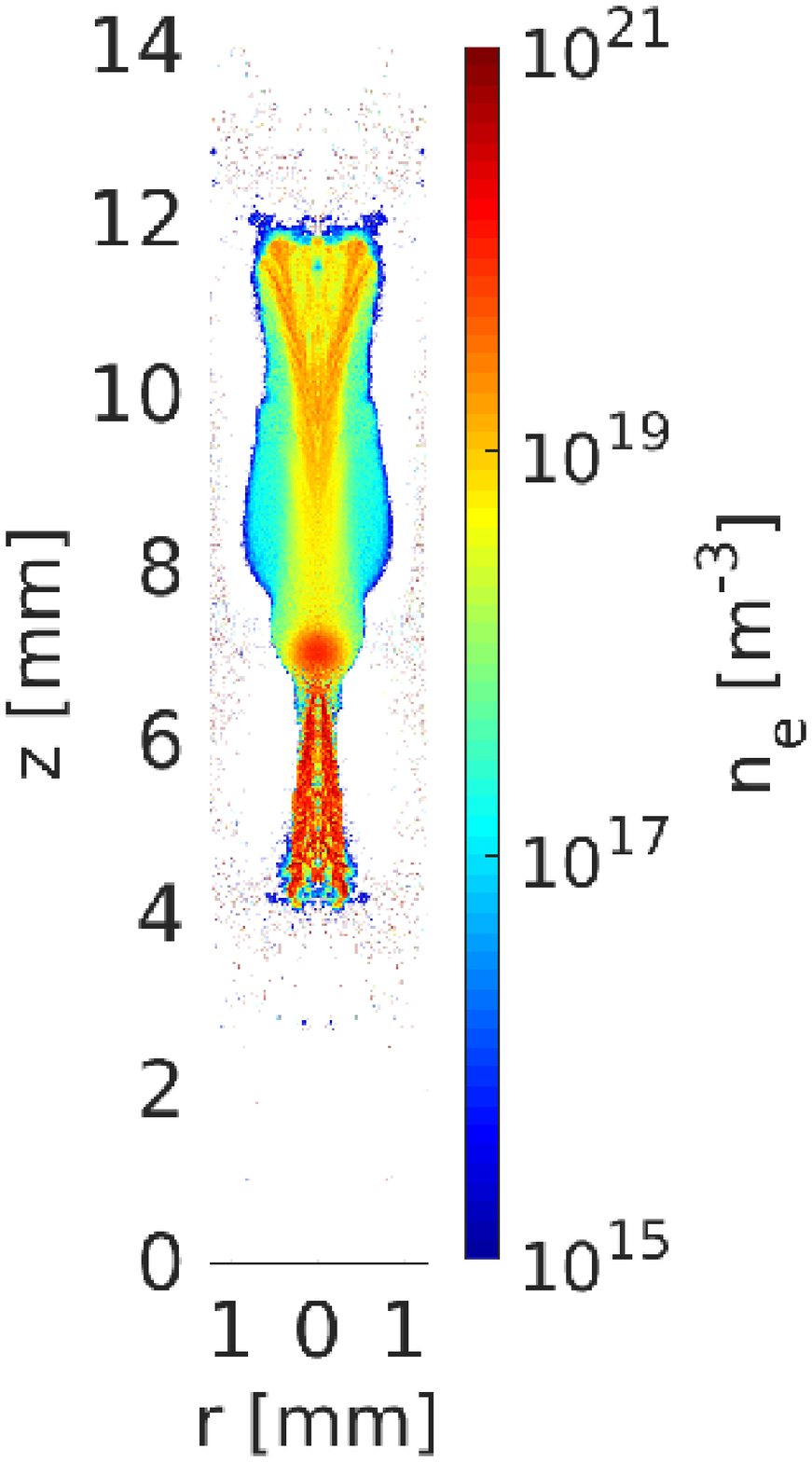} &
\includegraphics [scale=0.3] {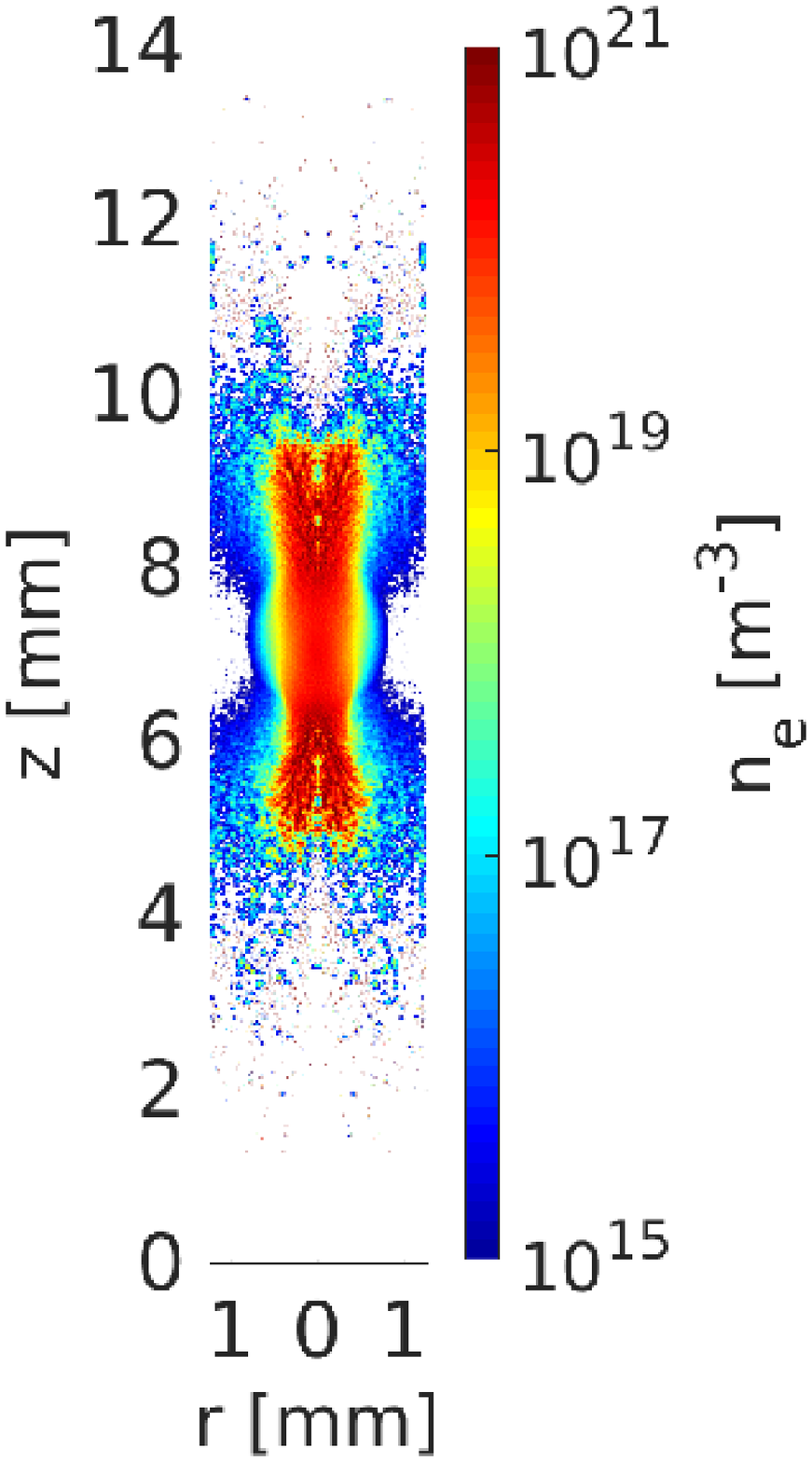} \\
\includegraphics [scale=0.3] {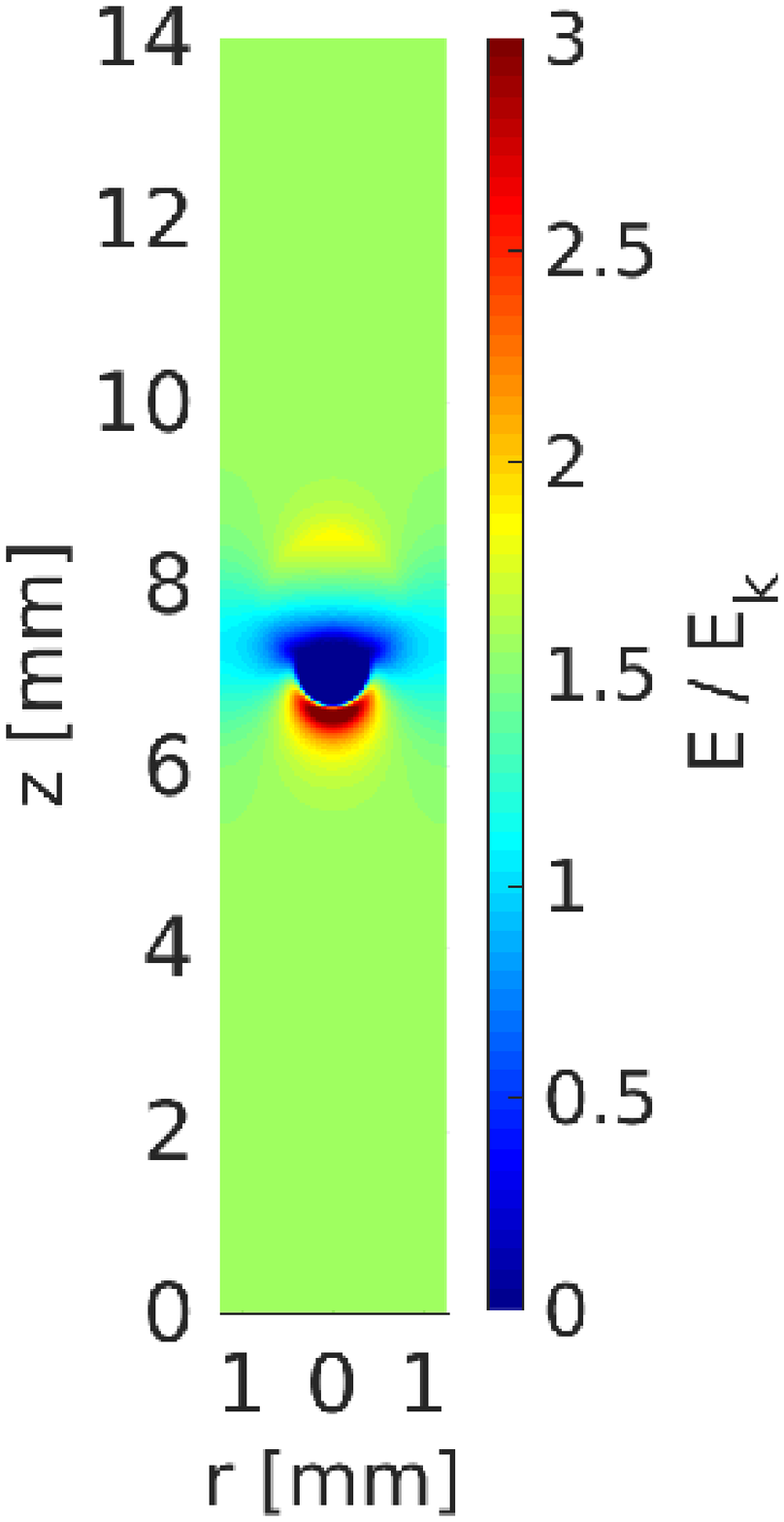} &
\includegraphics [scale=0.3] {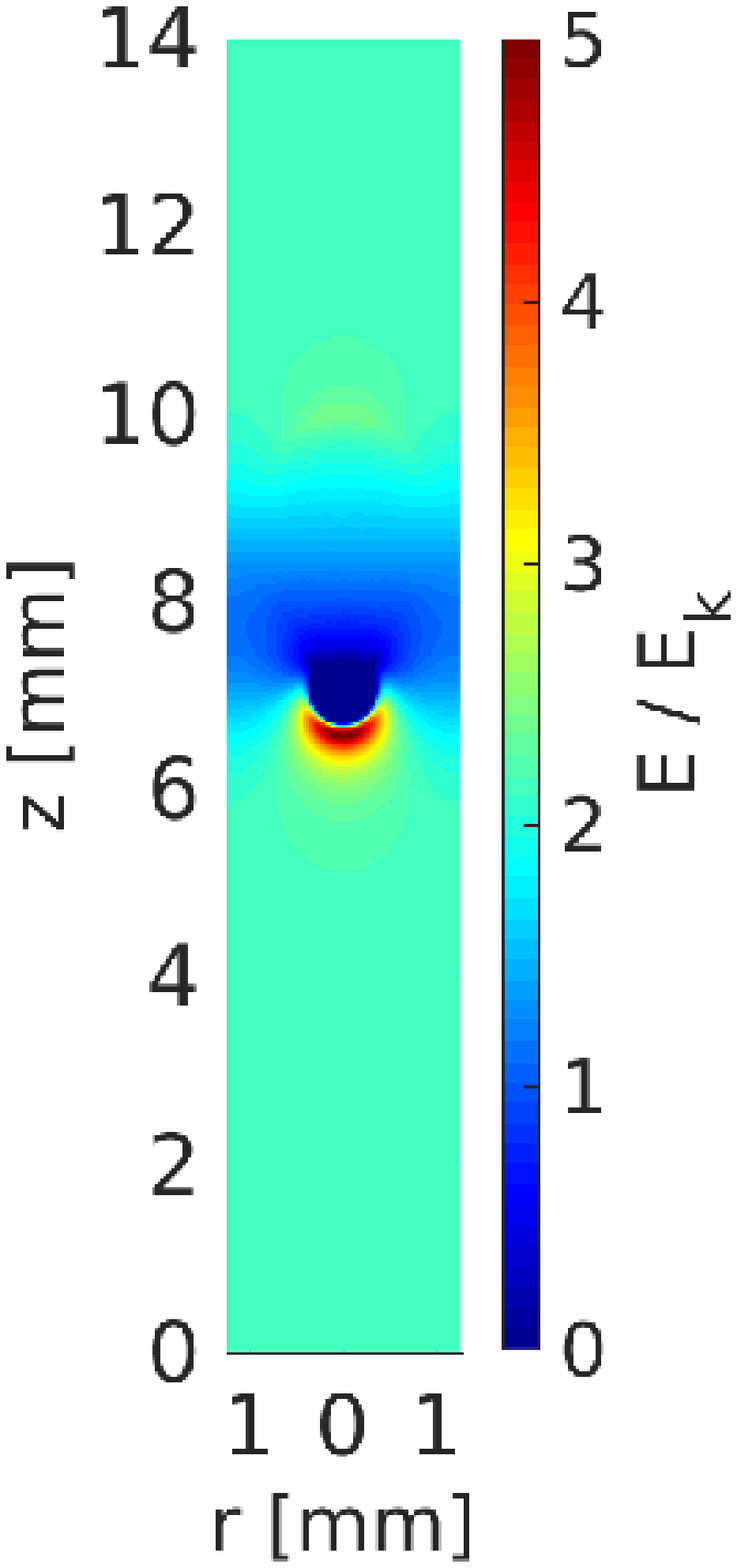} &
\includegraphics [scale=0.3] {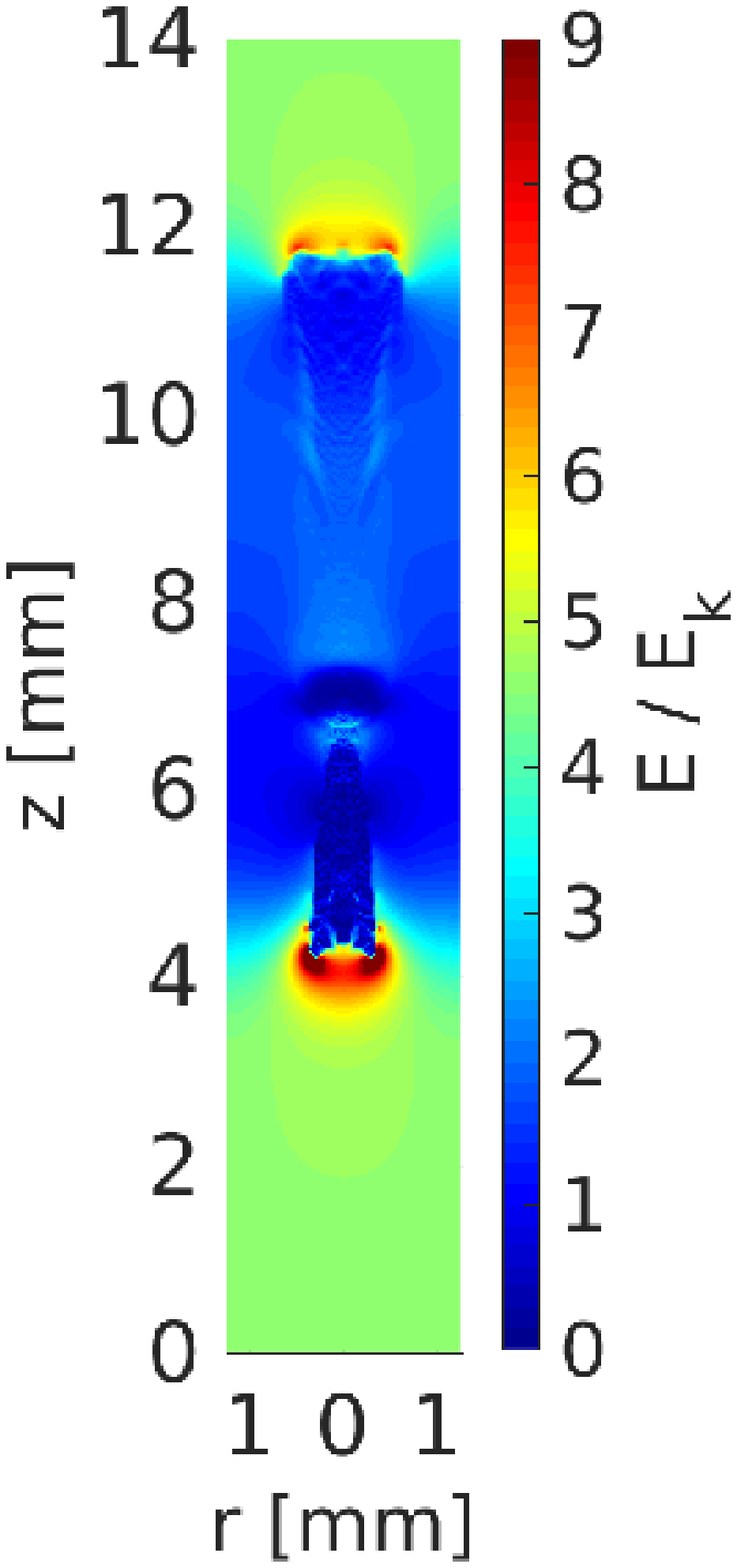} &
\includegraphics [scale=0.3] {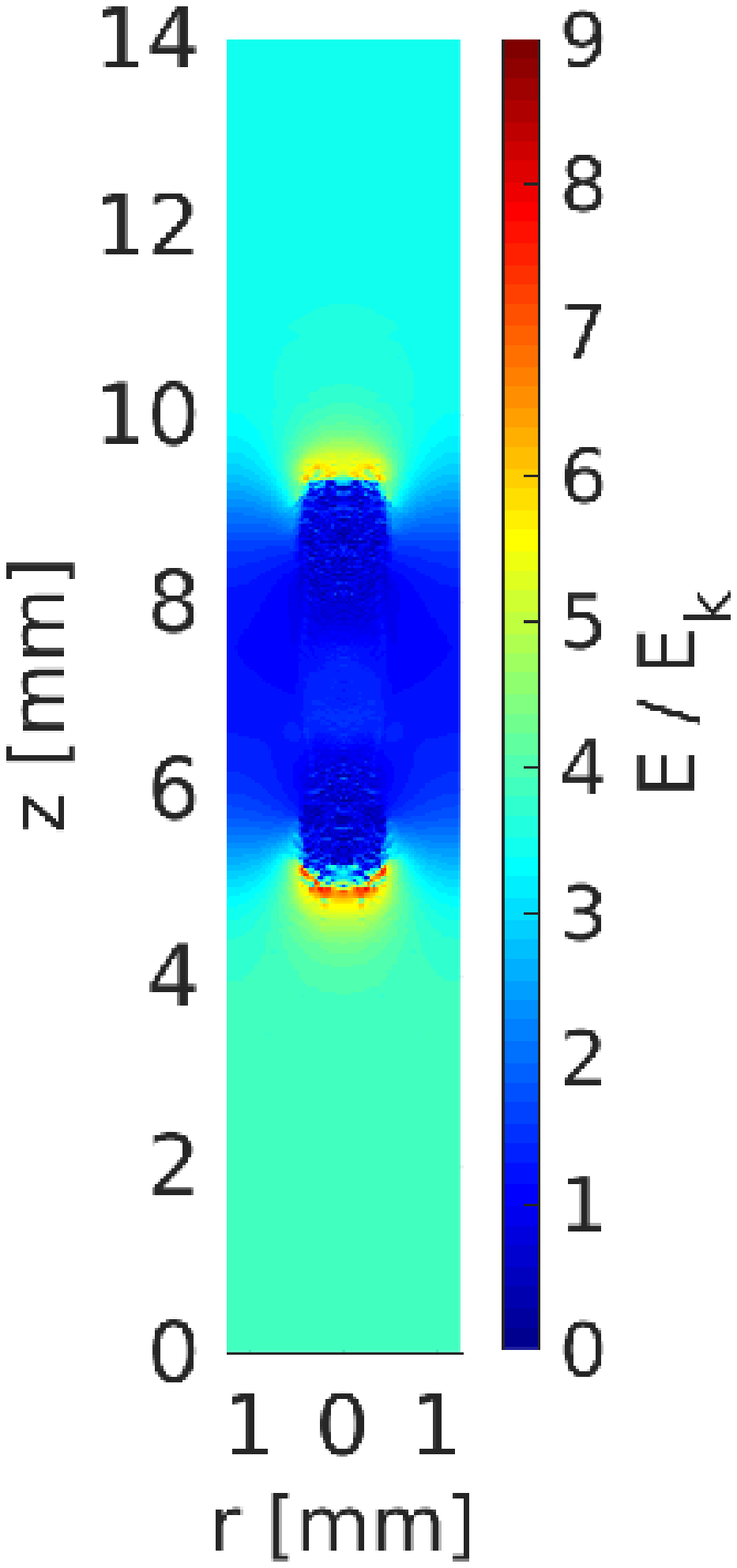} \\
$E=1.5E_k$ & $E=2E_k$ & $E=3E_k$ & $E=3E_k$\\
$t=9.33$ ns & $t=17.93$ ns & $t=10.39$ ns & $t=0.58$ ns\\
\end {tabular}
\end {center}
\caption {The electron density (first row) and the electric field (second
row) in N$_2$:CH$_4$=(98.4\%):(1.6\%) (columns one to three) and in N$_2$:O$_2$=
(98.4\%):(1.6\%) (fourth column) in different ambient fields after different time steps.} \label{dens_3.fig}
\end {figure}
%%%%%%%%%%%%%%%%%%%%%%%%%%%%%%%%%%%%%%%%%%%%%%%%%%%%%%%%%%%%%%%%%%%%%%

%%%%%%%%%%%%%%%%%%%%%%%FIG. 10%%%%%%%%%%%%%%%%%%%%%%%%%%%%%%%%%%%%%%%%
\begin {figure}
\includegraphics [scale=0.40] {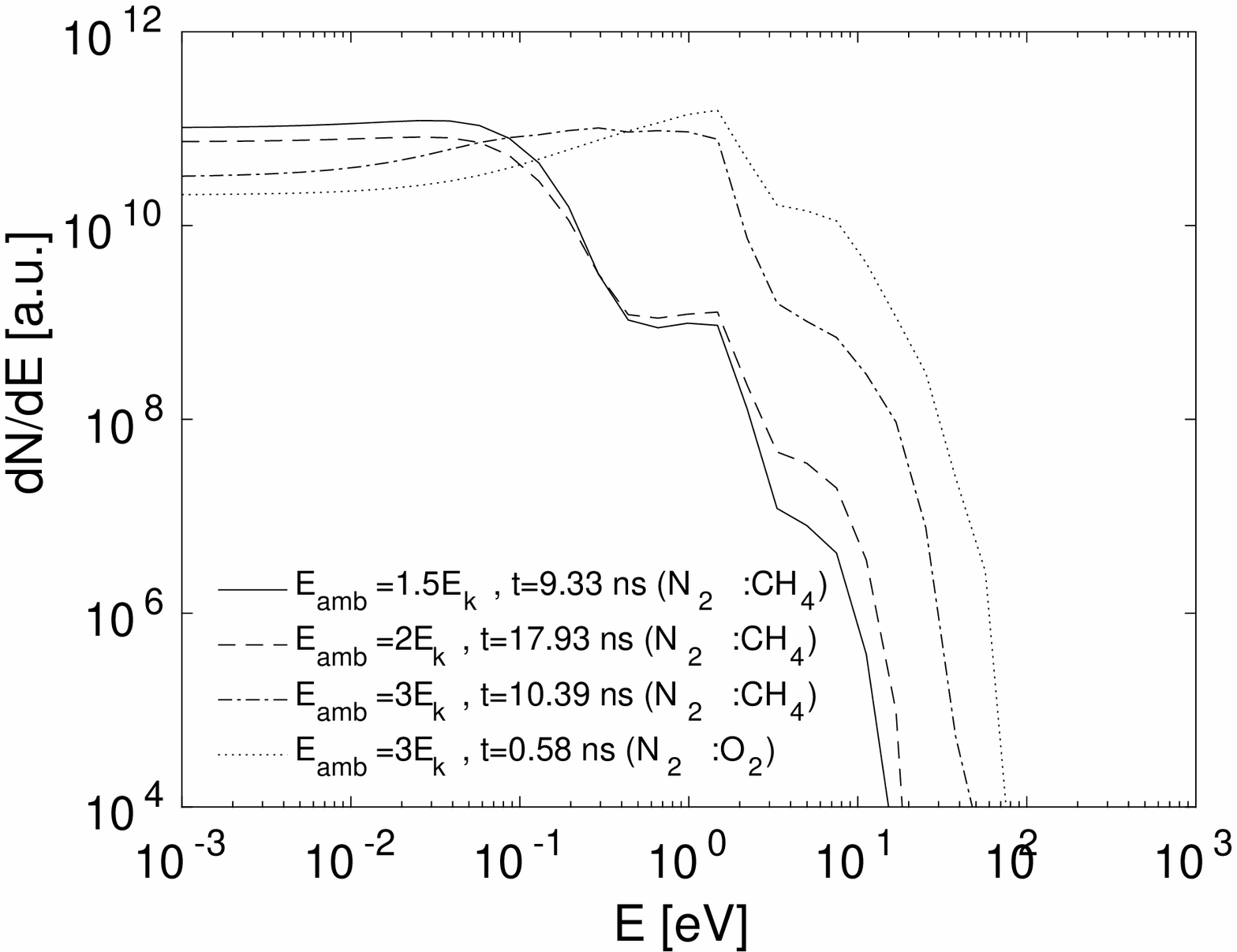}
\caption {The energy distribution for $\kappa=0.984$ for the same fields and
time steps as in Fig. \ref{dens_3.fig}. } \label{energy_2.fig}
\end {figure}
%%%%%%%%%%%%%%%%%%%%%%%%%%%%%%%%%%%%%%%%%%%%%%%%%%%%%%%%%%%%%%%%%%%%%%

After comparing the inception and evolution of bidirectional streamers
in different N$_2$:CH$_4$ and N$_2$:O$_2$ mixtures, we now focus on the
evolution of plasma patches and their transition to bidirectional
streamers on Saturn's moon Titan, hence in N$_2$:CH$_4$ with $98.4\%$ nitrogen. Figure \ref{dens_3.fig} shows the electron density and
the electric field in different ambient fields; we
here would like to remind the reader that the number density of ambient
molecules is $2.9\cdot 10^{25}$ m$^{-3}$ which corresponds to 20 km
altitude on Titan, i.e. typical cloud altitudes in its atmosphere.
As supplementary material, we have added movie files displaying the
temporal evolution of the electron density and the electric field for the cases presented in Fig. \ref{dens_3.fig}. For comparison, the last column shows the
electron density and the electric field in N$_2$:O$_2$ for the same number density of ambient gas molecules.
Similar to $\kappa=0.8$, we observe the
avalanche-to-streamer transition in N$_2$:CH$_4$ only for fields above $2E_k$.
As for $\kappa=0.8$, the ionization length in fields below $2E_k$ is longer in mixtures with
methane, thus the ionization is delayed and subsequently also the formation
of enhanced field tips. In
fields $\le 2E_k$, there is a distinct motion of electrons towards the positive electrode
only; there is no motion of the positive front at all. However, the field is
enhanced at the positive front and shielded at the negative front.
Figure \ref{energy_2.fig} shows the electron energies after the same time
steps and in the same electric fields as in Fig. \ref{dens_3.fig}. The maximum electron energy is approximately 20 eV for
fields smaller than $2E_k$ in N$_2$:CH$_4$, thus there is no efficient
drive of ionization.

For fields above $2E_k$, we observe an avalanche-to-streamer transition
within the borders of the simulation domain.
However, comparing the evolution on Titan (third column) and in N$_2$:O$_2$
(fourth column), reveals that both streamer fronts move significantly slower
in Titan's atmosphere. Additionally, the electron density in N$_2$:CH$_4$ is smaller and
branching is favored. The dot-dashed and dotted lines in Figure \ref{energy_2.fig} compare the electron
energies for these two cases. There is a significant
number of electrons above 20 eV resulting in more electron impact ionization. However, as
a consequence of the friction force, the number of electrons above 20 eV is amplified in N$_2$:O$_2$
in comparison to N$_2$:CH$_4$ which accelerates the formation of a double-headed streamer. Furthermore, the less probable photoionization in N$_2$:CH$_4$ additionally
delays the streamer development.

\section {Conclusions and outlook} \label {concl.sec}

%%%%%%%%%%%%%%%%%%%%%%%TAB. 1%%%%%%%%%%%%%%%%%%%%%%%%%%%%%%%%%%%%%%%%%
\begin {table}
\begin {tabular}{c|c|c|c}
\backslashbox{$\kappa$}{$E_{amb}$}& $1.5E_k$ & $2E_k$ & $3E_k$ \\
\hline
20\% & $\checkmark$ & $\checkmark$ & $\checkmark$\\
40\% & & $\checkmark$ & $\checkmark$\\
60\% & & $\checkmark$ & $\checkmark$\\
80\% & & & $\checkmark$\\
98.4\% & & & $\checkmark$
\end {tabular}
\caption {Criteria of successful avalanche-to-streamer transitions and subsequent
streamer inception within the simulation domain with $L_z=1.4$ cm in nitrogen-methane mixtures with different percentages $\kappa$ of nitrogen in different
ambient fields $E_{amb}$.} \label{trans.tab}
\end {table}
%%%%%%%%%%%%%%%%%%%%%%%%%%%%%%%%%%%%%%%%%%%%%%%%%%%%%%%%%%%%%%%%%%%%%%

We have investigated the motion of electrons and the streamer inception in
N$_2$:CH$_4$ and in N$_2$:O$_2$ mixtures with number densities of $2.9\cdot 10^{25}$ m$^{-3}$ with different percentages $\kappa$ of
nitrogen in ambient fields of $1.5E_k$, $2E_k$ and $3E_k$.
Whilst streamers form for all considered cases in N$_2$:O$_2$, we
observe the streamer inception in N$_2$:CH$_4$ mixtures depending on $\kappa$ and the ambient field $E_{amb}$. Table \ref{trans.tab} shows which combinations of
$\kappa$ and $E_{amb}$ favor the formation of
double-headed streamers in N$_2$:CH$_4$ within the simulation domain. There are two scenarios: For small
percentages of nitrogen, fields as low as $1.5E_k$ are sufficient to incept
streamers; if the percentage of nitrogen is
increased, higher and higher fields are required. Using a 1.5D fluid
model, we have observed the same tendency for negative streamers irrespective of the initial
electron density \cite{bosnjakovic_2018}. Ambient fields
slightly above the breakdown field, where the rate of ionization is higher than
the rate for attachment, the effective ionization coefficient for large $\kappa$ is
not sufficient to effectively ionize the ambient gas within the simulation
domain. For large percentages of nitrogen and small fields, the
ionization length can be larger than or comparable to the size of the simulation domain
preventing us from observing the inception of streamers.

Considering that for all
$\kappa$, the breakdown field in N$_2$:CH$_4$ is approximately half as large
as the breakdown field in N$_2$:O$_2$ and that the friction force below 3 eV is larger in N$_2$:CH$_4$,
fronts move one to two orders of magnitude faster in nitrogen-oxygen mixtures than in nitrogen-methane
mixtures independent of whether we observe an avalanche-to-streamer transition. As an additional effect, photoionization is less effective in
N$_2$:CH$_4$, hence, the motion of the fronts is further damped.

On Titan with methane percentages between 1.4\% and 5\%, we do not observe any streamer inception for ambient fields below $\approx 3E_k$ which
equals a field strength of 4.2 MV m$^{-1}$. Recent simulations
have calculated large scale electric fields as high as 2 MV m$^{-1}$ in
Titan's atmosphere \cite{tokano_2001}. Thus, at first glance it seems that streamer inception
is not feasible in Titan's
atmosphere. However, one may not forget that typical large scale electric
fields in Earth's thunderclouds are typically in the order of 0.1-0.2 MV m$^{-1}$
\cite{marshall_1995,marshall_2005} whereas the classical breakdown fields at
cloud altitudes vary between approximately 2.0 MV m$^{-1}$ at 4 km altitude and
0.5 MV m$^{-1}$ at 16 km altitude, thus approximately an order of magnitude larger. Although the origin of
lightning on Earth is still under debate \cite{gurevich_2013,dubinova_2015},
its existence, and thus also streamer inception, are very well observed despite the difference between the large-scale thundercloud fields and the breakdown field.
However, we note that the inception of streamers in Titan's
atmosphere strongly depends on the ambient field which is still
uncertain since it is only provided by models, but not by direct
measurements. Moreover, even if we conclude the existence of small
(cm-long) streamer discharges in Titan's atmosphere, this does not allow us to conclude about
the existence of lightning. Although their existence cannot be ruled out
completely, 127 flybys of Cassini did not reveal any traces of lightning
\cite {fischer_personal_communication}. Thus, either lightning was too weak
to be detected \cite{fischer_2011,fischer_personal_communication}, or it was
not present at all. Since even on Earth, the streamer-to-lightning-leader
transition is not fully understood yet \cite{silva_2013}, we propose further work
to study this transition in Titan's atmosphere.

Since the primordial atmosphere of Earth had a similar chemical
composition as Titan's atmosphere, our model here allows us to also study the
inception of streamers on the early Earth.

We here speculate that similar to the processes in terrestrial
streamer discharges,
run-away electrons and subsequently X-rays
might be produced in Titan's atmosphere. In future work, we will estimate the occurrence and the
fluence of these phenomena related to Titan $\gamma$-ray flashes (TGRFs) and discuss their effect on
Titan's atmosphere.

The model presented in this manuscript also allows to study the
streamer inception in different atmospheres, as for example, of exoplanets
\cite{bailey_2014,hodosan_2016a,hodosan_2016b}, provided appropriate cross sections for the
scattering of electrons off the atmospheres'
constituents are known. Hence, in future work, we strive to investigate which known
exo\-planets are likely to show discharge phenomena and also tackle the
question of associated high-energy beams.

\noindent
{\bf Acknowledgments:} We would like to thank Georg Fischer from the
Institut f\"ur Weltraumforschung (IWF), Graz, Austria, for fruitful discussions to improve the
paper. The research was partly funded by the Marie Curie
Actions of the European Union's Seventh Framework Programme (FP7/2007-2013)
under REA grant agreement n\textsuperscript{o} 609405
(COFUNDPostdocDTU). This project has received funding from the European
Union's Horizon 2020 research and innovation programme under the Marie
Sklodowska-Curie grant agreement 722337. SD acknowledges support from MPNTRRS Projects OI171037 and Ill41011.

\end {document}